\documentclass[printer]{tMOP2e}

\newcommand{\E}{\mathalpha{\mathrm{e}}}
\newcommand{\I}{\mathalpha{\mathrm{i}}}
\newcommand{\sds}[1]{\mbox{\small$\displaystyle{#1}$}}
\newcommand{\ds}{\displaystyle}
\newcommand{\ket}[1]{|#1\rangle}
\newcommand{\bra}[1]{\langle#1|}
\newcommand{\bvec}[1]{\boldsymbol{#1}}

\usepackage[english]{babel}
\usepackage{graphicx}
\usepackage[usenames,dvipsnames]{xcolor}

\makeatletter
\renewcommand{\draftnote}{\relax}
\renewcommand{\jname}{\relax}
\renewcommand{\jobtag}{\relax}
\def\articletype#1{\gdef\@articletype{#1}}
\gdef\@articletype{} 
\renewcommand{\@received}{Posted on the arXiv on 28 September 2012}
\renewcommand{\ps@plain}{
\renewcommand{\@oddfoot}{\hfil\thepage\hfil}\renewcommand{\@evenfoot}{}
\renewcommand{\@oddhead}{}\renewcommand{\@evenhead}{}}
\makeatother

\begin{document}

\markboth{Rui Han, Hui Khoon Ng and Berthold-Georg Englert}
{Raman transitions without adiabatic elimination}

\title{Raman transitions without adiabatic elimination: \\
A simple and accurate treatment}

\author{Rui HAN$^{a}$, Hui Khoon NG$^{a,b}$ and  Berthold-Georg ENGLERT$^{a,c}$
\\\vspace{6pt}  
$^{a}$\textit{Centre for Quantum Technologies, %
National University of Singapore, Singapore 117543, Singapore}\\
$^{b}$\textit{DSO National Laboratories, 20 Science Park Drive, %
Singapore 118230, Singapore}\\
$^{c}$\textit{Department of Physics, National University of Singapore, %
Singapore 117542, Singapore}\\
}

\date{\today}
\maketitle  

\begin{abstract}
Driven Raman processes --- nearly resonant two-photon transitions through an
intermediate state that is non-resonantly coupled and does not acquire a
sizeable population --- are commonly treated with a simplified description in
which the intermediate state is removed by adiabatic elimination. 
While the adiabatic-elimination approximation is reliable when the detuning of
the intermediate state is quite large, it cannot be trusted in other
situations, and it does not allow one to estimate the population in the
eliminated state. 
We introduce an alternative method that keeps all states in the description,
without increasing the complexity by much. An integro-differential equation of
Lippmann-Schwinger type generates a hierarchy of approximations, but very
accurate results are already obtained in the lowest order.  
\end{abstract}


\section{Introduction}
Atomic and molecular systems can be coupled in various ways, so that the
atomic states evolve and the populations of states change. 
Of all the electromagnetic multipole couplings, the electric dipole is the
strongest. Thus optically allowed atomic transitions are often stimulated by
an electric-dipole coupling between the atom and a well-controlled laser.

Dipole-allowed transitions can be driven with optical lasers directly. 
It is, however, quite common that the desired transition is dipole forbidden
or the transition frequency is outside of the popular optical range. Then
some intermediate state can assist in an indirect transition. 
Examples for such processes include three-level Raman
transitions~\cite{ScullyZubairy:97}, 
multi-level Raman transitions~\cite{PhysRevA.35.4207, Deng1983284}, and
stimulated Raman adiabatic passage 
(STIRAP)~\cite{PhysRevA.29.690, PhysRevA.40.6741}. 
They all have their useful applications, with three-level Raman transitions
being perhaps the most widely used in a large variety of experiments.

In order to reduce the computational complexity of dealing with large Hilbert
spaces, one can decrease the dimensionality of the system by eliminating
states that are not populated much or not coupled strongly. 
For a typical three-level Raman transition, the intermediate state is far
off-resonantly coupled to the relevant initial and target states. This enables
one 
to perform the so-called \emph{adiabatic elimination} that gets rid of the less
relevant state and yields a two-level effective Hamiltonian. 
Although the procedure of adiabatic elimination  is well understood 
--- see \cite{Brion+2:07} and \cite{Paulisch+3:12}, for instance --- it
gives a reliable approximation only when the detuning of the intermediate state
is much larger than the Rabi frequencies for the coupling to the other 
two states.

In this paper, we introduce an alternative approach to the quantitative
description of the three-level Raman transition. 
This new method does not rely on adiabatic elimination and gives a much more
accurate solution while the computational complexity remains low. 
We set the stage in Section~\ref{sec:3levels}, where we briefly review driven
three-level systems and state the notational conventions used throughout. 
Then, Section~\ref{sec:AE} deals with the usual adiabatic-elimination approach
and comments on its limitations and problems. 
Our new approach is explained in Section~\ref{sec:New}: First we present 
the general methodology, then we show how it gives the exact solution to the
resonant two-photon transition problem, and finally we generate reliable
approximations for the situation of a non-resonant two-photon transition. 
We close with a summary and outlook.

\section{The three-level system}\label{sec:3levels}
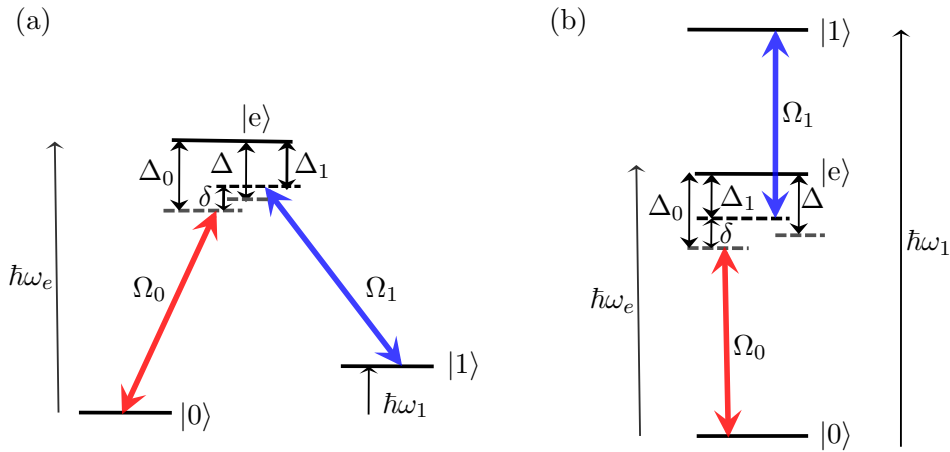
\begin{figure}[h]
\centerline{\setlength{\unitlength}{0.88pt}
\begin{picture}(400,220)(0,0)
\put(0,10){\includegraphics[scale=0.7]{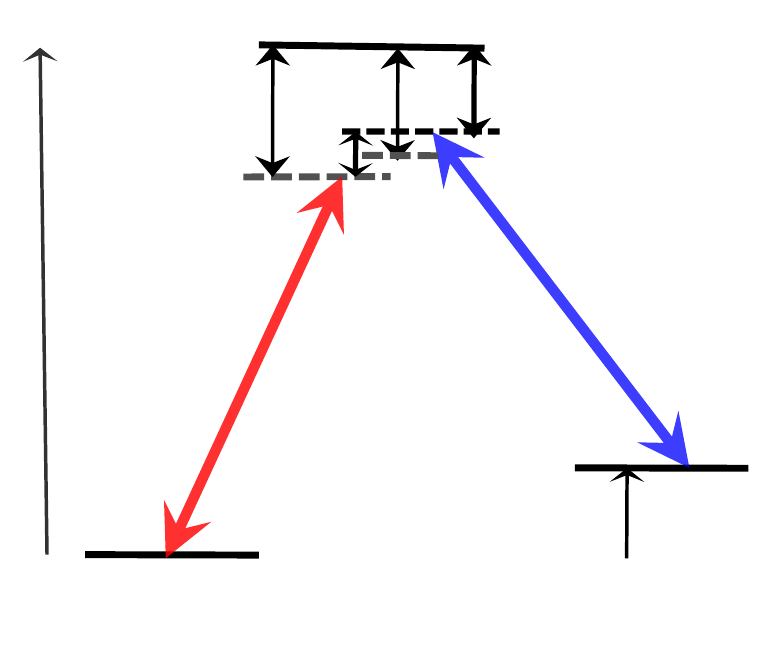}\label{ramanlevellambda}}
\put(250,0){\includegraphics[scale=0.7]{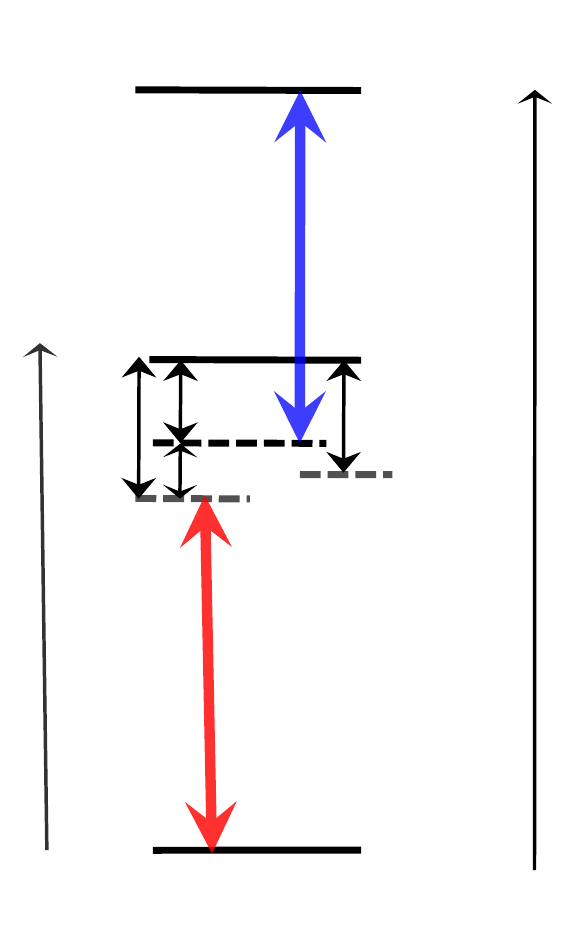}\label{ramanlevelcas}}
\put(0,200){\makebox(0,0){(a)}}
\put(-1,90){\makebox(0,0){$\hbar\omega_e$}}
\put(160,35){\makebox(0,0){$\hbar\omega_1$}}
\put(70,30){\makebox(0,0){$\ket{0}$}}
\put(185,52){\makebox(0,0){$\ket{1}$}}
\put(96,158){\makebox(0,0){$\ket{\mathrm{e}}$}}
\put(50,85){\makebox(0,0){$\Omega_0$}}
\put(150,85){\makebox(0,0){$\Omega_1$}}
\put(53,135){\makebox(0,0){$\Delta_0$}}
\put(120,138){\makebox(0,0){$\Delta_1$}}
\put(81,139){\makebox(0,0){$\Delta$}}
\put(74,125){\makebox(0,0){$\delta$}}
\put(230,200){\makebox(0,0){(b)}}
\put(249,80){\makebox(0,0){$\hbar\omega_e$}}
\put(385,105){\makebox(0,0){$\hbar\omega_1$}}
\put(345,22){\makebox(0,0){$\ket{0}$}}
\put(345,195){\makebox(0,0){$\ket{1}$}}
\put(345,135){\makebox(0,0){$\ket{\mathrm{e}}$}}
\put(308,60){\makebox(0,0){$\Omega_0$}}
\put(329,160){\makebox(0,0){$\Omega_1$}}
\put(272,120){\makebox(0,0){$\Delta_0$}}
\put(303,124){\makebox(0,0){$\Delta_1$}}
\put(335,124){\makebox(0,0){$\Delta$}}
\put(298,109){\makebox(0,0){$\delta$}}
\end{picture}}
\caption{Level scheme of a typical Raman transition. (a) shows the level
  structure of a $\Lambda$-type Raman transition and (b) shows the level
  structure of a cascade-type Raman transition. $\Omega_0$ and $\Omega_1$
  denote the Rabi frequencies of the individual two-level transitions,
  $\Delta$ denotes the detuning of the laser from the transition frequency of
  the excited state and $\delta$ is the detuning of the two-photon
  transition. The requirement is that the detuning $\Delta$ is much larger
  than the Rabi frequencies so that the excited state $\ket{\mathrm{e}}$ is
  not significantly populated.}\label{fig:raman} 
\end{figure}

A Raman transition is a two-photon process that gives an
effective coupling between two states $\ket{0}$ and $\ket{1}$ via a
far-detuned auxiliary state $\ket{\mathrm{e}}$; see Fig.~\ref{fig:raman}.  
As mentioned above, Raman transitions are often used when the transition
between levels $\ket{0}$ and $\ket{1}$ is dipole forbidden or has an
inconvenient frequency. 
The $\Lambda$-type configuration of Fig.~\ref{fig:raman}(a) applies to
transitions between different ground states via an excited state; 
the cascade-type configuration of Fig.~\ref{fig:raman}(b) can be used to
achieve the transition between a ground state and a highly excited state, such
as a Ryd\-berg state.
There is also a V-type transition with the level configuration similar to that
of an inverted $\Lambda$-type transition.  
Although the level structures are different for these configurations, 
the underlying physics is essentially the same. 

We will, therefore, restrict ourselves to treating the $\Lambda$-type Raman
transition in detail.  
With reference to $\ket{0}$, $\ket{1}$, and $\ket{\mathrm{e}}$ in this order,
the $3\times3$ matrix for the Hamiltonian of the system is  
\begin{equation}\label{eq:RamanHS}
H=H_{\mathrm{Atom}}+H_{\mathrm{AL}}\,,
\end{equation}
where 
\begin{equation}\label{eq:HAtom}
H_{\mathrm{Atom}}=\hbar{\left(
\begin{array}{ccc}0&0&0\\0&\omega_1&0\\0&0&\omega_{\mathrm{e}}\end{array}\right)}
\end{equation}
is the part for the atom by itself with the convention that the state
$\ket{0}$ has energy zero, and 
\begin{equation}\label{eq:HAL}
H_{\mathrm{AL}}=\frac{\hbar}{2}{\left(
\begin{array}{ccc}0&0&\Omega_0 \E^{\I\omega_{\mathrm{L0}}t}\\
                  0&0&\Omega_1 \E^{\I\omega_{\mathrm{L1}}t}\\
                  \Omega_0^* \E^{-\I\omega_{\mathrm{L0}}t}&\Omega_1^*
                  \E^{-\I\omega_{\mathrm{L1}}t}&0
\end{array}\right)}
\end{equation}
accounts for the atom-laser interaction. 
The laser frequencies are denoted by $\omega_{\mathrm{L0}}$ and
$\omega_{\mathrm{L1}}$, respectively. 
The Rabi frequencies for the electric-dipole transitions are 
\begin{eqnarray}
\Omega_0=\frac{q_{\mathrm{el}}}{\hbar}\bra{0}
         \bvec{r}\cdot\bvec{E}_{\mathrm{L0}}\ket{\mathrm{e}}
\quad\mathrm{and}\quad
\Omega_1=\frac{q_{\mathrm{el}}}{\hbar}\bra{1}
         \bvec{r}\cdot\bvec{E}_{\mathrm{L1}}\ket{\mathrm{e}}\,,
\end{eqnarray}
where $q_{\mathrm{el}}$ is the electron charge, $\bvec{E}_{\mathrm{L0}}$ and
$\bvec{E}_{\mathrm{L1}}$ are the amplitudes of the electric fields of the
laser beams and $\bvec{r}$ is the position vector of the atom. 
Each Rabi frequency depends on the intensity and polarization of the
corresponding laser as well as the dipole matrix element between the two
coupled states, and can be complex. 
Typically, the detuning $\Delta$ is designed to be large, i.e., 
$|\Delta|\gg |\Omega_0|,|\Omega_1|$, so that the auxiliary state
$\ket{\mathrm{e}}$ does not get populated to avoid problems arising from
uncontrolled spontaneous emission from $\ket{\mathrm{e}}$ to other states; see
Fig.~\ref{fig:raman}.
Since the two-photon transition from $\ket{0}$ to $\ket{1}$ is nearly
resonant, the overall detuning $\delta$ of the two-photon transition is small
compared with the average detuning $\Delta$, $|\delta/\Delta|\ll1$. 
If the detunings between the lasers and the atomic frequencies are denoted by
$\Delta_0=\omega_{\mathrm{e}}-\omega_{\mathrm{L0}}$ and
$\Delta_1=\omega_{\mathrm{e}}-\omega_1-\omega_{\mathrm{L1}}$, we
have $\Delta=(\Delta_0+\Delta_1)/2$ and $\delta=\Delta_0-\Delta_1$.

In view of the time-dependent phase factors in $H_{\mathrm{AL}}$, it is
expedient to switch to an interaction picture in which the Hamiltonian does
not depend on time.
This is achieved by identifying the ``free'' Hamiltonian $H_0$ by a suitable
splitting of the atomic Hamiltonian,
\begin{equation}\label{RamanH0}
H_0=H_\mathrm{Atom}+\hbar{\left(
\begin{array}{c@{\ }cc}\frac{1}{2}\delta&0&0\\
                       \ 0&-\frac{1}{2}\delta&0\\
                       \ 0&0&-\Delta\end{array}\right)}
=\hbar{\left(
\begin{array}{c@{\ }cc}\frac{1}{2}\delta&0&0\\
                       \ 0&\omega_1-\frac{1}{2}\delta&0\\
                       \ 0&0&\omega_{\mathrm{e}}-\Delta\end{array}\right)},
\end{equation}
for which we obtain the interaction-picture Hamiltonian
\begin{equation}\label{RamanHI}
H_{\mathrm{I}}=\E^{\I H_0t/\hbar}(H-H_0) \E^{-\I H_0t/\hbar}
=\frac{\hbar}{2}{\left(\begin{array}{ccc}-\delta &0& \Omega_0\\
0 & \delta & \Omega_1 \\
\Omega_0^*&\Omega_1^* & 2\Delta\end{array}\right)}.
\end{equation}
The evolution of the three-level system is then studied with the aid of the
Schr\"odinger equation
\begin{equation}\label{eq:SchrEq}
  \I\hbar\frac{\partial}{\partial t}{\Psi}_{\mathrm{I}}(t)
   =H_{\mathrm{I}}\Psi_{\mathrm{I}}(t)\qquad\mbox{with}
\quad\Psi_{\mathrm{I}}(t)=\left(
\begin{array}{c}c_0(t)\\c_1(t)\\c_{\mathrm{e}}(t)\end{array}
\right)=\E^{\I H_0t/\hbar}\Psi(t)\,,
\end{equation}
where the components of $\Psi_{\mathrm{I}}(t)$ are the interaction-picture
probability amplitudes for $\ket{0}$, $\ket{1}$, and $\ket{\mathrm{e}}$,
related to the respective components of $\Psi(t)$, the amplitudes in the
Schr\"odinger picture, by simple time-dependent phase factors. 
As a consequence of this simple relation between the components of $\Psi(t)$
and  
$\Psi_{\mathrm{I}}(t)$, we can simply square $c_0(t)$, $c_1(t)$, or
$c_{\mathrm{e}}(t)$ to obtain the probability amplitudes for the respective
atomic levels, as exemplified by the probability for $\ket{0}$,
\begin{equation}
  \Bigl|\bigl(\begin{array}{ccc}1&0&0\end{array}\bigr)\Psi(t)\Bigr|^2
=\Bigl|\bigl(\begin{array}{ccc}1&0&0\end{array}\bigr)\Psi_{\mathrm{I}}(t)\Bigr|^2
=\bigl|c_0(t)\bigr|^2\,,
\end{equation}
where, of course, $\bigl(\begin{array}{ccc}1&0&0\end{array}\bigr)$ is the
three-component row for $\bra{0}$.

To understand the system analytically, we can solve for the eigensystem of
this time-independent Hamiltonian directly. 
However, very often, this is neither the most efficient way of getting the
solution, nor the best method for obtaining a good physical insight into the
system. 
Various approaches have been developed, of which the adiabatic
elimination and, for $\delta=0$, the dark-state method~\cite{PhysRevA.54.794} 
are particularly useful and popular.

\section{Adiabatic elimination}\label{sec:AE}
\subsection{The methodology}\label{ssec:AE1}
The standard textbook approach to the Raman-transition problem makes use of
``adiabatic elimination'' in accordance with the following line of reasoning. 
Since the excited state $\ket{\mathrm{e}}$ is far-detuned by $\Delta$, it will
remain barely populated if it has no initial population.
Thus, the change of the population in this state can be taken as approximately
zero, $\mbox{\small$\ds\frac{\partial}{\partial t}$}c_\mathrm{e}(t)=0$, so that
\begin{equation}\label{changece}
2\I\frac{\partial}{\partial t}c_\mathrm{e}(t)
=\Omega_0^*c_0(t)+\Omega_1^*c_1(t)+2\Delta c_{\mathrm{e}}(t)=0,
\end{equation}
in view of the Schr\"odinger equation (\ref{eq:SchrEq}).
We can now express $c_{\mathrm{e}}(t)$ as a linear combination of $c_0(t)$ and
$c_1(t)$, and so eliminate $c_{\mathrm{e}}(t)$ from the equations of motion for
$c_0(t)$ and $c_1(t)$.
This gives us an effective $2\times2$ Hamiltonian for the evolution of
the two relevant states 
\begin{equation}\label{eq:HeffRaman}
H_{\mathrm{eff}}=-\frac{\hbar}{2}{\left(\begin{array}{c@{\ }c} 
\delta+\frac{|\Omega_0|^2}{2\Delta} & \frac{\Omega_0\Omega_1^*}{2\Delta} 
\\[1ex]
\frac{\Omega_1\Omega_0^*}{2\Delta} & -\delta+\frac{|\Omega_1|^2}{2\Delta}
\end{array}\right)}
\quad\mbox{for}\quad\I\hbar\frac{\partial}{\partial t}{\psi}(t)
   =H_{\mathrm{eff}}\psi(t)\quad\mbox{with}
\quad\psi(t)={\left(
\begin{array}{c}c_0(t)\\c_1(t)\end{array}
\right)}.
\end{equation}
As an immediate benefit of applying adiabatic elimination on the intermediate
auxiliary state, the effective Hamiltonian is a simple $2\times2$ matrix, 
for which the eigenvalues and the projectors to the eigenspaces are readily
available.

The eigenvalues of $H_{\mathrm{eff}}$ are
\begin{equation}
E_\pm=-\frac{\hbar}{8\Delta}\bigl(|\Omega_0|^2+|\Omega_1|^2\bigr)
\pm\frac{\hbar}{2} \Omega_{\mathrm{R}}
\end{equation}
with the positive frequency $\Omega_{\mathrm{R}}$ given by
\begin{equation}\label{RamanFreEff}
\Omega_{\mathrm{R}}^2=\frac{1}{(4\Delta)^2}\bigl(|\Omega_0|^2+|\Omega_1|^2\bigr)^2
+\frac{\delta}{2\Delta}\bigl(|\Omega_0|^2-|\Omega_1|^2\bigr)+\delta^2.
\end{equation}
The projectors to the corresponding eigenspaces are
$\frac{1}{2}(1\pm\sigma_o)$, where
\begin{equation}
\sigma_o=\frac{2H_\mathrm{eff}-E_+-E_-}{E_+-E_-}
\end{equation}
is a Pauli-type matrix. 
If the evolution starts with all the population in the ground state $\ket{0}$, 
i.e., $c_0(t=0)=1$, the population in state $\ket{1}$ at a later time $t$ 
is
\begin{equation}\label{c1t0}
\bigl|c_1(t)\bigr|^2=\frac{|\Omega_0|^2|\Omega_1|^2}{8\Delta^2\Omega_{\mathrm{R}}^2}
\bigl[1-\cos(\Omega_{\mathrm{R}}t)\bigr]\,.
\end{equation}
This tells us the physical significance of $\Omega_{\mathrm{R}}=(E_+-E_-)/\hbar$: 
It is the effective Rabi frequency of the transition between states $\ket{0}$
and $\ket{1}$ via this Raman process. 

In the situation of vanishing overall detuning, $\delta=0$, we have
\begin{equation}\label{c1t1}
\bigl|c_1(t)\bigr|^2=\frac{2|\Omega_0|^2|\Omega_1|^2}
{\bigl(|\Omega_0|^2+|\Omega_1|^2\bigr)^2}
{\left[1-\cos{\left(\frac{|\Omega_0|^2+|\Omega_1|^2}{4\Delta}t\right)}\right]}\,,
\end{equation}
where the effective Rabi frequency is 
$\Omega_{\mathrm{R}}=\bigl(|\Omega_0|^2+|\Omega_1|^2\bigr)/(4|\Delta|)$ and the
amplitude of the Rabi oscillation is less than unity,
\begin{equation}
  \frac{4|\Omega_0|^2|\Omega_1|^2}{\bigl(|\Omega_0|^2+|\Omega_1|^2\bigr)^2}<1\,,
\end{equation}
unless $|\Omega_0|^2=|\Omega_1|^2$. 
In other words, when $\delta=0$, we get complete population transfer between
state $\ket{0}$ and state $\ket{1}$ only if the two lasers drive the
respective transitions equally strongly. 

More generally, perfect population transfer from $\ket{0}$ to $\ket{1}$
is only possible if the two diagonal matrix elements of the effective
Hamiltonian are identical, i.e., if there is no effective detuning after the
adiabatic elimination.
Then $\ket{0}$ and $\ket{1}$ are equal-weight superpositions of the
eigenstates of $H_{\mathrm{eff}}$ and temporal evolution turns one into the other. 
For given laser intensities, and thus given Rabi frequencies $\Omega_0$ and
$\Omega_1$, the experimenter can exploit the Zeeman or the Stark effect to
adjust the overall detuning such that  
\begin{equation}\label{delta1}
\delta=\frac{|\Omega_1|^2-|\Omega_0|^2}{4\Delta}\,.
\end{equation}
This makes the effective detuning vanish and ensures perfect population
transfer. 
Indeed, for this value of $\delta$, the right-hand side of (\ref{c1t0})
simplifies, 
\begin{equation}\label{c1t2}
\bigl|c_1(t)\bigr|^2
=\frac{1}{2}{\left[1-\cos\left(\Omega_{\mathrm{R}}t\right)\right]}
\quad\mbox{with}\quad\Omega_{\mathrm{R}}=\frac{|\Omega_0||\Omega_1|}{2|\Delta|}\,.
\end{equation}
In experiments, specifically for two-photon population transfer from the ground
state to a Ryd\-berg state where the Rabi frequencies $|\Omega_0|$ and
$|\Omega_1|$ can be an order of magnitude different in strengths, adjusting
the detuning in accordance with (\ref{delta1}) is
important~\cite{PhysRevA.82.013405}.

\subsection{Light shift}\label{ssec:AE2}
When an atomic transition is driven by an electromagnetic radiation field with
detuning $\Delta$, the dressed atomic levels are shifted. 
This is the so-called \emph{light shift} of the atomic levels, which is a
second-order 
correction to the eigenenergies of the Hamiltonian~\cite{BarrattCohen:61}. 
If an atomic transition is driven by Rabi frequency $\Omega$ with detuning
$\Delta$, the light shift of the ground state of the transition is
$-|\Omega|^2/(4\Delta)$ and the light shift of the dressed excited state is of
the same amount but opposite in sign.
The overall light shift is a direct summation of the light shifts arising from
individual electromagnetic radiation fields, when the atomic level is addressed
by multiple fields. 

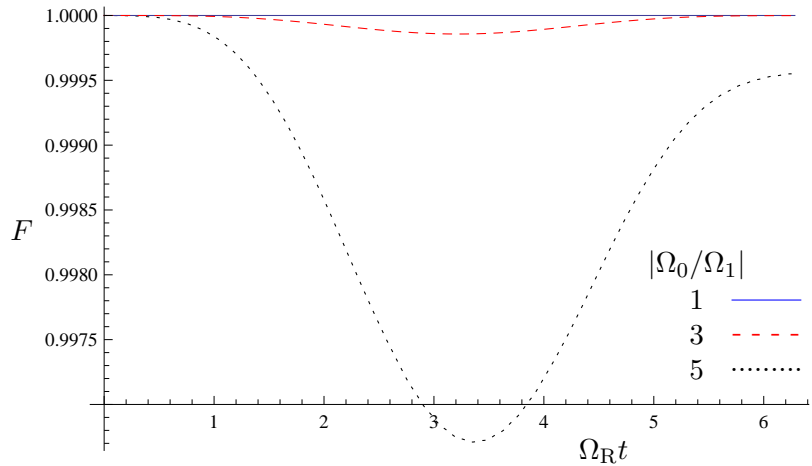
\begin{figure}
\centerline{\setlength{\unitlength}{0.88pt}
\begin{picture}(345,205)(-125,-7)
\put(-110,-0){\includegraphics[scale=0.8]{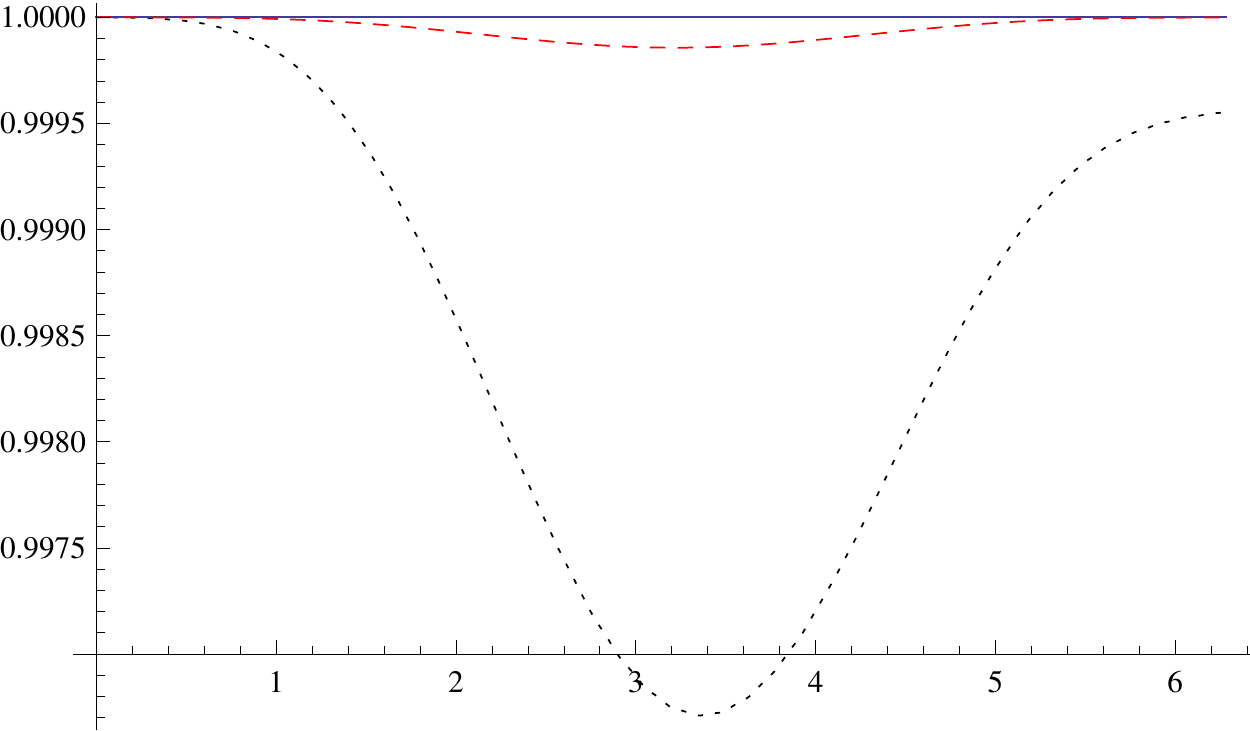}}
\put(-120,95){\makebox(0,0){$F$}}
\put(130,0){\makebox(0,0){$\Omega_\mathrm{R}t$}}
\put(70,0){
\put(100,80){\makebox(0,0){$|\Omega_0/\Omega_1|$}}
\put(100,65){\makebox(0,0){1}}
\put(100,50){\makebox(0,0){3}}
\put(100,35){\makebox(0,0){5}}
\put(130,65){\makebox(0,0){\color{blue}\line(1,0){30}}}
\put(130,50){\makebox(0,0){\color{red}- - - -}}
\put(130,35){\makebox(0,0){\color{black}.........}}
}
\end{picture}}
\caption{\label{fig:F}
Fidelity between the two states at later time evolving with the exact
Hamiltonian but for the two different $\delta$ values of 
(\ref{delta1}) and (\ref{delta2}). 
$\Delta=400$Mhz and $|\Omega_1|=40$Mhz are fixed and the different curves
are for three different values of the ratio $|\Omega_0/\Omega_1|$.}
\end{figure}

In the present context, the light shift of state $\ket{0}$ is
$\hbar\delta_0=-\hbar|\Omega_0|^2/(4\Delta_0)$ and the light shift of state
$\ket{1}$ is $\hbar\delta_1=-\hbar|\Omega_1|^2/(4\Delta_1)$. 
Thus, the value of $\delta$ that brings the two-photon transitions into
resonance is determined by
\begin{equation}
\delta=\frac{|\Omega_1|^2}{4\Delta+2\delta}
       -\frac{|\Omega_0|^2}{4\Delta-2\delta}\,,
\end{equation}
which is solved by
\begin{equation}\label{delta2}
\delta\simeq \frac{2\Delta\bigl(|\Omega_1|^2-|\Omega_0|^2\bigr)}
                  {8\Delta^2+|\Omega_0|^2+|\Omega_1|^2}+\cdots\,,
\end{equation}
where the ellipsis stands for terms of relative size $|\Omega/\Delta|^2$ or
smaller. 
The difference between the $\delta$ values obtained from (\ref{delta1})
and (\ref{delta2}) can be of the order of a percent 
($\sim(|\Omega_0|^2+|\Omega_1|^2)/(64\Delta^2)$). 
But the approximate solutions provided by the adiabatic-elimination method
do not depend much on this small fractional difference in $\delta$. 
Figure~\ref{fig:F} shows the fidelity between the states at later times using
the two different values of $\delta$ when the initial state is
$\ket{0}$. 

The effect of using these two different $\delta$ values is small when
the difference between $|\Omega_0|$ and $|\Omega_1|$ is not too large. 
For the cases shown in Fig.~\ref{fig:F}, the error from using (\ref{delta1})
is a small fraction of a percent. 
This explains why (\ref{delta1}) can serve as a good guidance for experiments,
although (\ref{delta2}) is more accurate.

\subsection{Problems with adiabatic elimination}\label{ssec:AE3}
In physics, one speaks of an ``adiabatic process'' if a relevant property
evolves quite slowly, whereas other processes are fast --- a clear separation of
time scales is a defining element. 
In the context of adiabatic elimination, one invokes such a separation in the
evolution of the components of $\Psi_{\mathrm{I}}(t)$, for which we have the
Schr\"odinger equation in (\ref{eq:SchrEq}). 
One standard argument observes that, as a consequence of
$|\Delta|\gg|\Omega_0|,|\Omega_1|$, the amplitude $c_{\mathrm{e}}(t)$ will
undergo many oscillations during a period in which $c_0(t)$ and $c_1(t)$ do
not change substantially. 
Then, if our interest is not in the short-time changes that would reveal the
rapid oscillations of $c_{\mathrm{e}}(t)$ but predominantly in the longer-time
changes of $c_0(t)$ and $c_1(t)$, the average change of $c_{\mathrm{e}}(t)$
over these longer periods is expected to be quite small. 
In the spirit of this reasoning, we should then read (\ref{changece}) as a
statement about coarse-grained values of the probability amplitudes.     

This is hardly a rigorous argument, and whether one regards it as convincing
or not is largely a matter of taste.
Clearly, though, a more solid argument would be welcome, and one has been
provided in~\cite{Paulisch+3:12}.  
Indeed, the reasoning in~\cite{Paulisch+3:12} uses coarse graining in
conjunction with a Markov approximation.    

There are other problems that one needs to keep in mind when employing
adiabatic elimination.
We present them as four questions.
\begin{enumerate}\renewcommand{\theenumi}{(\roman{enumi})}
\item Which is the correct interaction picture to use? 
  The theory argues that the change of the (coarse-grained) population in the
  excited state is approximately zero and we arrived at (\ref{changece}) by
  using the Schr\"odinger's equation of motion with the time-independent
  Hamiltonian $H_{\mathrm{I}}$ in (\ref{RamanHI}). 
  However, the choice of interaction picture is not unique and why do we apply
  the adiabatic elimination in this particular interaction picture instead of
  another? 
  For example, by adding a constant term, the operator $H_{\mathrm{I}}+E$
  describes the system equally well.
  But if we use an interaction picture with $E\ne0$, the resultant two-level
  effective Hamilton operator also changes as $\Delta$ gets replaced by
  $\Delta+E/\hbar$ in (\ref{eq:HeffRaman}) with consequential changes in the
  evolution of $\psi(t)$.

  This ambiguity in the choice of interaction picture was studied by Brion,
  Pedersen, and M{\o}lmer with a Green's function formalism~\cite{Brion+2:07}.
  It would be premature, however, to claim that the case is closed. Further
  studies of the choice of interaction picture and a systematic way of
  performing the adiabatic elimination are presented in a companion
  paper~\cite{Paulisch+3:12}. 

\item Is it possible to estimate the population in the eliminated state?
  This does not have a simple answer because there are problems with the
  normalization of the wave function. 
  Originally, we have $\Psi_{\mathrm{I}}(t)^{\dagger}\Psi_{\mathrm{I}}(t)^{\ }=1$.
  Then, since $H_\mathrm{eff}$ is hermitian, we also have
  $\psi(t)^{\dagger}\psi(t)=1$. 
  Combined with the initial condition $(c_0,c_1,c_{\mathrm{e}})=(1,0,0)$ 
  this implies ${c_\mathrm{e}(t)= 0}$ for all $t$, which contradicts the basic
  approximation in (\ref{changece}).
  
\item Is it possible to regard the adiabatic-elimination approximation as the
  first in a hierarchy of approximations?
  This is possible, indeed, as discussed in \cite{Paulisch+3:12}.
  It turns out that the next approximation in the hierarchy gives a
  substantial improvement and better quantitative estimates. 
  In addition, it answers the previous question inasmuch as the
  next approximation provides an estimate for the population in the auxiliary
  state. 

\item Is it possible to avoid the adiabatic elimination without increasing the
  complexity much beyond the convenient two-level description of
  (\ref{eq:HeffRaman})? 
  Yes, this is possible, as we demonstrate in Section~\ref{sec:New}.
\end{enumerate}

\section{Without adiabatic elimination }\label{sec:New}
Alternative methods other than adiabatic elimination are also used to solve
Raman-transition problems. 
The most direct way is, of course, to diagonalize the interaction-picture
Hamiltonian of (\ref{RamanHI}),
and this can be done by hand since the dimensionality of the system is
small. 
But the expressions for the eigenvalues and eigencolumns of $H_{\mathrm{I}}$
are quite involved and not transparent.  
An exception is the two-photon transition with vanishing overall detuning, 
${\delta=0}$, when one can
identify a dark state and use it to reduce the three-level system to an
effective two-level system, which can then be solved exactly rather simply. 
The dark states are particularly useful in the context of adiabatic population
transfer and electromagnetically induced transparency. 
When the overall detuning is nonzero, however, there is no dark state. 
Methods of perturbation theory can then be used to find corrections for a small
detuning $\delta$, but the complexity grows quickly when high accuracy is
required.  

In this section, we provide a new way of solving the Raman-transition
problem. 
Just like the dark-state method, it gives a compact exact solution for
${\delta=0}$, and it can solve the ${\delta\neq0}$ case
with very high precision and rather little extra effort.

\subsection{General methodology}\label{ssec:New1}
It will be expedient to use a different interaction picture as the one of
(\ref{RamanH0}) and (\ref{RamanHI}) that we used for the 
adiabatic elimination in Section~\ref{ssec:AE1}. 
Instead, we choose
\begin{equation}
H_0=\frac{\hbar}{2}{\left(\begin{array}{ccc}\Delta+\delta &0& 0\\
0 & 2\omega_1+\Delta-\delta &0 \\
0&0 & 2\omega_{\mathrm{e}}-\Delta\end{array}\right)}\,,
\end{equation}
and the Hamiltonian in this interaction picture is 
\begin{equation}\label{RamanH1,Inew}
H_{\mathrm{I}}=\frac{\hbar}{2}
{\left(\begin{array}{ccc}-\Delta-\delta &0& \Omega_0\\
0 & -\Delta+\delta & \Omega_1 \\
\Omega_0^*&\Omega_1^* & \Delta\end{array}\right)}
=\frac{\hbar}{2}
{\left(\begin{array}{cc}-(\Delta+\delta\sigma_3)&\Omega\\
\Omega^{\dagger}&{\Delta}\end{array}\right)}\,,
\end{equation}
where the latter way of writing emphasizes the ${3=2+1}$ split into two 
relevant states and one auxiliary state; $\sigma_3$ is the standard third
Pauli $2\times2$ matrix, and $\Omega$ and $\Omega^{\dagger}$ are the
two-component column of Rabi frequencies, 
\begin{equation}
\Omega={\left(\begin{array}{c}\Omega_0\\\Omega_1\end{array}\right)}\,,
\end{equation}
and its adjoint row.
This change of interaction picture is equivalent to shifting the energy levels
of the interaction-picture Hamiltonian in (\ref{RamanHI}) by $-\hbar\Delta/2$,
an example of the freedom of choice discussed in question (i) in Section
\ref{ssec:AE3}.  

We note that the square of $H_{\mathrm{I}}$, 
\begin{equation}\label{eq:HI2}
H_{\mathrm{I}}^2=(\hbar M)^2=\hbar^2(M_0^2+\epsilon)\,, 
\end{equation}
is the sum of a ``big'' block-diagonal part,
\begin{equation}\label{eq:HI02}
{M_0}^2 \equiv \frac{1}{4}
{\left(\begin{array}{cc} (\Delta+\delta\sigma_3)^2+\Omega \Omega^{\dagger}&0
\\0&\Delta^2+\Omega^{\dagger} \Omega\end{array}\right)}\,,
\end{equation}
and a ``small'' off-diagonal part
\begin{equation}\label{eq:H_1}
\epsilon \equiv -\frac{\delta}{2}{\left(\begin{array}{cc} 0&\sigma_3 \Omega\\
                      \Omega^{\dagger}\sigma_3&0\end{array}\right)}\,.
\end{equation}
Matrix $M$ denotes a square root of the matrix $(H_{\mathrm{I}}/\hbar)^2$ and,
since for each eigenvalue we have a choice of sign, there are many $M$s that
are equally good. 
Any one can be used as a replacement of $H_{\mathrm{I}}/\hbar$ in even
functions of $H_{\mathrm{I}}$; for example, we could choose $M>0$ by convention.
In particular, then, the unitary evolution matrix in this interaction picture 
can be written as
\begin{equation}\label{eq:EvoR}
U(t)=\E^{-\I H_{\mathrm{I}}t/\hbar}=\cos(Mt)
-\frac{\I}{\hbar}\frac{\sin(Mt)}{M}H_{\mathrm{I}}\,.
\end{equation}
Except for the common physical approximations that enter the modeling of the
atom-laser system by a driven three-level system described by the Hamiltonian 
of (\ref{eq:RamanHS})--(\ref{eq:HAL}), this is an exact 
$3\times3$ matrix representing the evolution operator. 

Instead of diagonalizing the interaction-picture Hamiltonian
(\ref{RamanH1,Inew}), we can determine the eigenvalues and eigencolumns 
of $M^2$, whose ``big plus small'' structure, together with the block-diagonal
form of $M_0^2$, facilitates approximations. 
We will see the advantage thereof shortly. 
Let us note that we can position the factor $H_{\mathrm{I}}$ in the second
term of the right-hand-side of (\ref{eq:EvoR}) equally well to the left of
$\sin(Mt)/M$, or break up $\sin(Mt)/M$ and sandwich $H_{\mathrm{I}}$
between even powers of $M$. 
Since $[H_{\mathrm{I}},M^2]=0$, such a change in the order of the matrices makes
no difference in (\ref{eq:EvoR}), but slightly different expressions are
obtained when approximations are introduced for the trigonometric functions of
$M$.

\subsection{Vanishing overall detuning ($\delta=0$) %
--- exact solution}\label{ssec:New2}

When $\delta=0$, we have $\epsilon=0$ and
\begin{equation}\label{M_00}
M=M_0=\frac{1}{2}{\left(\begin{array}{cc}-\sqrt{\Delta^2+\Omega\Omega^{\dagger}}
&0\\0&\sqrt{\Delta^2+\Omega^{\dagger}\Omega}\end{array}\right)}\,,
\end{equation}
where the signs are chosen such that 
$M\to H_{\mathrm{I}}/\hbar$ in the $\Omega\to0$ limit.
The resulting evolution operator (\ref{eq:EvoR}) reads 
\begin{equation}\label{eq:EvoR0}
U(t)=\cos(M_0t)-\frac{\I}{\hbar}\frac{\sin(M_0t)}{M_0}H_{\mathrm{I}}\,.
\end{equation}
Owing to the block-diagonal structure of $M_0$, the original $3\times3$
problem has been converted into an equivalent $2\times2$ problem without
introducing any approximation. 
Clearly, the technical difficulty has been significantly reduced! 

The column $\Omega$ is an eigencolumn of $\Delta^2+\Omega\Omega^{\dagger}$
with eigenvalue $\Delta^2+\Omega^{\dagger}\Omega$, so that
\begin{equation}\label{eq:spectralM0}
  M_0^2=\frac{1}{4}(\Delta^2+\Omega^{\dagger}\Omega){\left(
      \begin{array}{cc}
      \sds{\frac{\Omega\Omega^{\dagger}}{\Omega^{\dagger}\Omega}} & 0\\[1ex]
       0 & 1        
      \end{array}\right)}
      +\frac{1}{4}\Delta^2
       {\left(\begin{array}{cc}
       1-\sds{\frac{\Omega\Omega^{\dagger}}{\Omega^{\dagger}\Omega}}& 0\\[1ex]
       0 & 0\end{array}\right)}
\end{equation}
is the spectral decomposition of $M_0^2$. 
This gives
\begin{equation}\label{eq:spectralCos}
  \cos(M_0t)=\cos\Bigl(\sqrt{\Delta^2+\Omega^{\dagger}\Omega}\,t/2\Bigr)
      {\left(\begin{array}{cc}
      \sds{\frac{\Omega\Omega^{\dagger}}{\Omega^{\dagger}\Omega}} & 0\\[1ex]
       0 & 1        
      \end{array}\right)}
      +\cos(\Delta t/2)
       {\left(\begin{array}{cc}
       1-\sds{\frac{\Omega\Omega^{\dagger}}{\Omega^{\dagger}\Omega}}& 0\\[1ex]
       0 & 0\end{array}\right)}
\end{equation}
and likewise for $\sin(M_0t)$ in (\ref{eq:EvoR0}).
Hence, the exact evolution of the system can be written out analytically. 
In particular, the population in the excited state $\ket{\mathrm{e}}$ is
non-zero.
With the atom initially in the ground state $\ket{0}$, it is 
\begin{equation}\label{eq:ce-exact}
\bigl|c_{\mathrm{e}}(t)\bigr|^2={\left|
(\begin{array}{ccc}0&0&1\end{array})\,U(t)
{\left(\begin{array}{c}1\\0\\0\end{array}\right)}\right|}^2
=\frac{|\Omega_0|^2}{\Delta^2+\Omega^{\dagger}\Omega}
\sin\Bigl(\sqrt{\Delta^2+\Omega^{\dagger}\Omega}\,t/2\Bigr)^2\,.
\end{equation} 
This exact expression shows that the population in the excited state
oscillates with (angular) frequency
$\frac{1}{2}(\Delta^2+\Omega^{\dagger}\Omega)^{1/2}$ and the
oscillation amplitude can be non-negligible if $|\Omega_0|^2$ is a sizeable
fraction of $\Delta^2+\Omega^{\dagger}\Omega=\Delta^2+|\Omega_0|^2+|\Omega_1|^2$. 

We note that the second projection matrix in (\ref{eq:spectralM0}) and
(\ref{eq:spectralCos}) projects on the dark state, whose bra has the row
${(-\Omega_1\ \Omega_0\ 0)}$~\cite{PhysRevA.54.794}.
An atom prepared in this dark state stays in it, and there is no
probability of finding the atom in the excited state at any time.
The atom is essentially decoupled from the driving lasers under these
circumstances.  
In this sense, one could regard (\ref{eq:EvoR0}) with (\ref{eq:spectralM0})
as the evolution matrix in the dark-state formalism but this is, in fact, not 
the case.
In the dark-state approach, one diagonalizes $H_{\mathrm{I}}$, reduced
to a two-dimensional problem after putting the dark state aside, which amounts
to choosing one particular square root of $M_0^2$ from the continuous family
of square roots that the degenerate eigenvalue makes available, namely the
square root whose eigenvalues and eigencolumns are those of
$H_\mathrm{I}/\hbar$. 
No such unique $M_0$ is needed in (\ref{eq:EvoR0}), nor is there any benefit
from enforcing a unique square root of $M_0^2$ by imposing additional criteria.
Although in the $\delta=0$ case, this equivalence can be established between
the new approach and the dark-state approach, the new approach offers more
flexibility and it also has a clear advantage when dealing with the
$\delta\ne0$ case, as we shall see in Section~\ref{ssec:New3}. 

\begin{figure}
\center{
\begin{picture}(450,300)(-30,-10)
\put(-25,150){\includegraphics[scale=0.7]{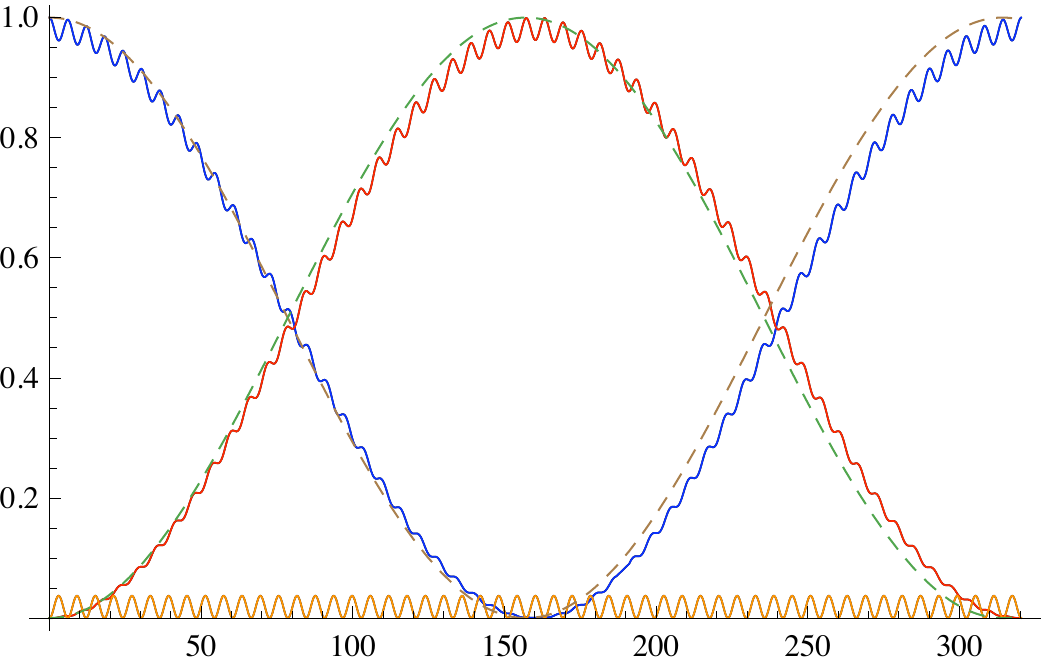}}
\put(155,150){\makebox(0,0){\small$\Delta t$}}
\put(200,150){\includegraphics[scale=0.7]{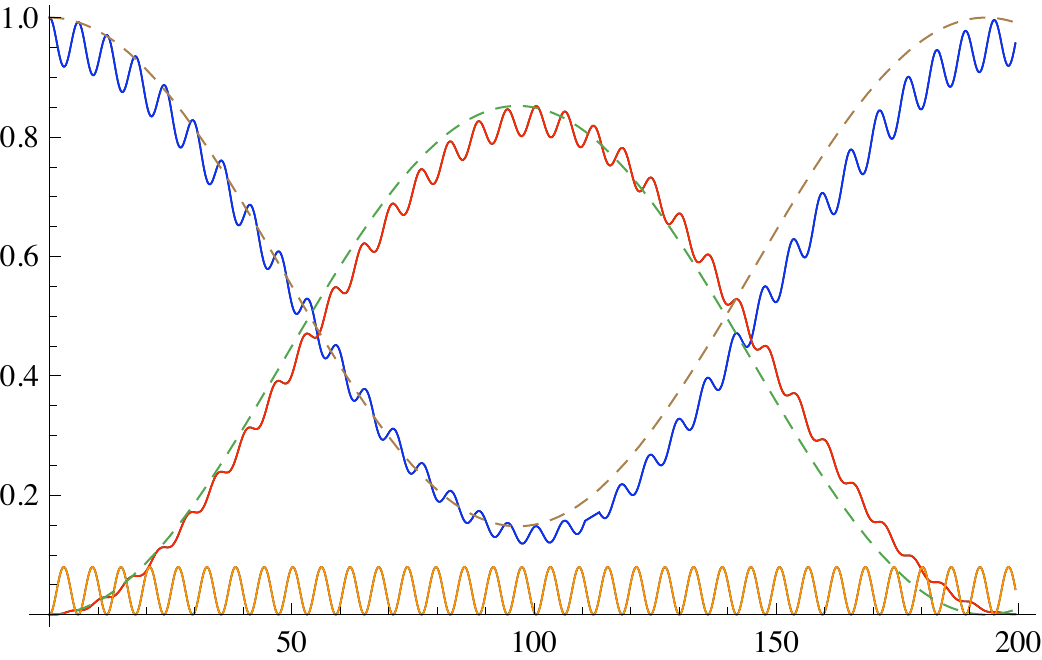}}
\put(380,150){\makebox(0,0){\small$\Delta t$}}
\put(-25,0){\includegraphics[scale=0.7]{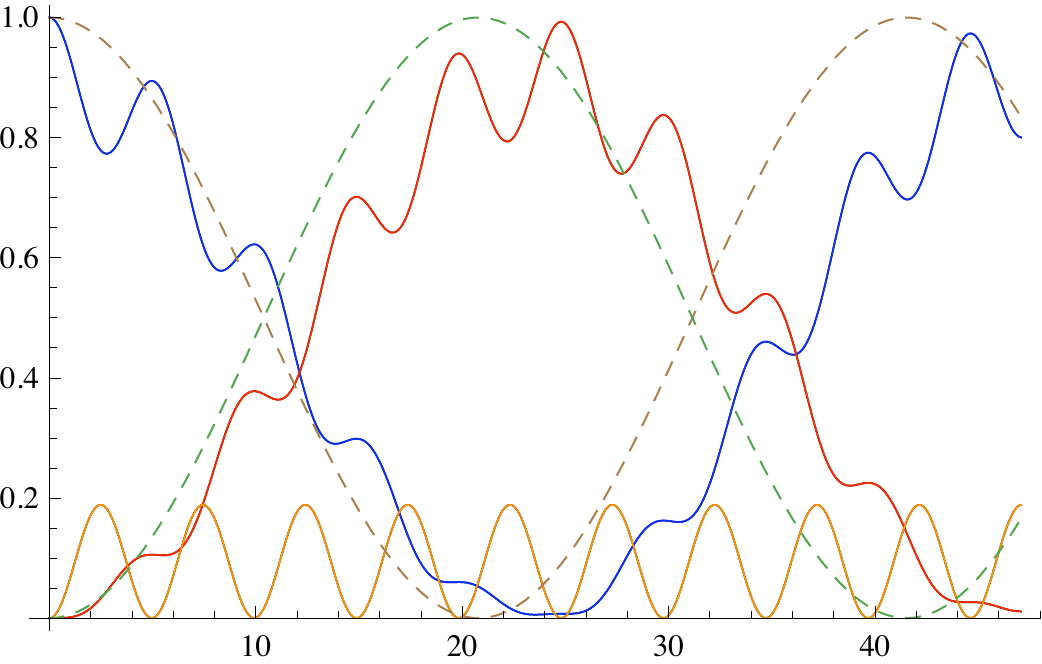}}
\put(130,0){\makebox(0,0){\small$\Delta t$}}
\put(220,120){\linethickness{0.8pt}\color{MidnightBlue}\line(1,0){20}}
\put(270,120){\makebox(0,0){$\;|c_0|^2$, exact}}
\put(220,105){\linethickness{0.8pt}\color{red}\line(1,0){20}}
\put(270,105){\makebox(0,0){$\;|c_1|^2$, exact}}
\put(220,90){\linethickness{0.8pt}\color{BurntOrange}\line(1,0){20}}
\put(270,90){\makebox(0,0){$\;|c_\mathrm{e}|^2$, exact}}
\put(230,75){\makebox(0,0){\color{Brown}{- - -}}}
\put(310,75){\makebox(0,0){$|c_0|^2$, adiabatic elimination}}
\put(230,60){\makebox(0,0){\color{green}{- - -}}}
\put(310,60){\makebox(0,0){$|c_1|^2$, adiabatic elimination}}
\end{picture}}
\caption{\label{fig:RamanOnRe}
Population distribution for a single-atom Raman transition in time when
$\delta=0$. 
The solid curves show the exact solution, and the dashed curves the
adiabatic-elimination approximation. 
The initial state is $\ket{0}$. 
The red curves are for the ground-state population $|c_0(t)|^2$, the
blue curves for $|c_1(t)|^2$, and the orange curves report
$|c_\mathrm{e}(t)|^2$, the population in the excited state. 
The detuning is $\Delta=400\,\mathrm{MHz}$ for all plots; 
the top left plot is for $|\Omega_0|=|\Omega_1|=\Delta/10=40\,\mathrm{MHz}$; 
the top right plot is for $|\Omega_0|=\Delta/10=40\,\mathrm{MHz}$
and $|\Omega_1|=\Delta/16=25\,\mathrm{MHz}$;
the bottom plot is for $|\Omega_0|=|\Omega_1|=\Delta/4=100\,\mathrm{MHz}$.}
\end{figure}
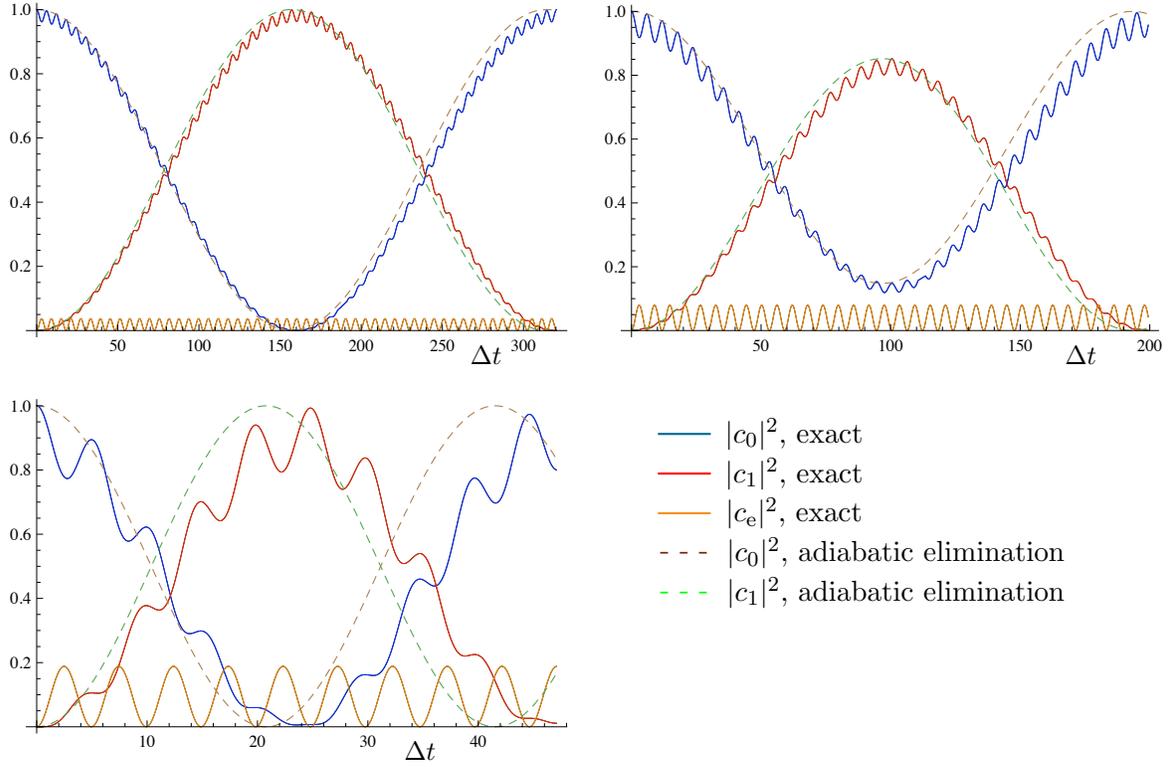

For three sets of parameter values, Fig.~\ref{fig:RamanOnRe} shows the
population $|c_{\mathrm{e}}(t)|^2$ of (\ref{eq:ce-exact}) as well as the
populations $|c_0(t)|^2$ and $|c_1(t)|^2$ of the relevant states $\ket{0}$ and
$\ket{1}$. 
The populations of $\ket{0}$ and $\ket{1}$ are oscillating with a slow
frequency that goes with a large amplitude and a fast frequency that goes with
a small amplitude. 
The population transfer is predominantly controlled by the slow frequency which,
therefore, is the effective Rabi frequency of the system. 
Since $M_0^2$ has only two distinct eigenvalues, half the 
sum of their square roots gives the fast frequency and the difference 
gives the slow frequency. 
Thus, we have
\begin{equation}\label{eq:Omega0R0}
\Omega_{\mathrm{R}}=\frac{1}{2}\Bigl(\sqrt{\Delta^2+\Omega^{\dagger}\Omega}
-|\Delta|\Bigr)=\frac{\Omega^{\dagger}\Omega}{4|\Delta|}
-\frac{(\Omega^{\dagger}\Omega)^2}{16|\Delta|^3}+\cdots
\end{equation}
for the effective Rabi frequency.
The expansion in powers of $|\Omega/\Delta|\ll1$ permits a comparison of this
exact expression with the approximation obtained by adiabatic elimination, the
$\delta=0$ version of (\ref{RamanFreEff}): 
The adiabatic-elimination approximation gives the leading term, but does not
reproduce any of the higher-order terms.

In summary, we find that the solution from the adiabatic elimination is indeed
the zeroth-order approximation of the exact result in the expansion of 
$\Omega^{\dagger}\Omega/\Delta^2$.
For example, in the zeroth order of $\Omega^{\dagger}\Omega/\Delta^2$, 
the excited state population vanishes as the oscillation amplitude 
is proportional to $\Omega^{\dagger}\Omega/\Delta^2\sim0$. 
We conclude that adiabatic elimination yields a reliable approximation only
when $|\Omega_0|,|\Omega_1|\ll|\Delta|$.
The difference between the solutions obtained by our method and the
adiabatic-elimination approximation 
is demonstrated clearly in Fig.~\ref{fig:RamanOnRe}.

The top left plot in Fig.~\ref{fig:RamanOnRe} shows the populations when
$|\Omega_0|=|\Omega_1|=\Delta/10$ is small; in this parameter regime, we can
already see the deviation of the adiabatic elimination from the exact result
but the deviation is not significant. 
In the top-right plot, the population of state $\ket{1}$ only reaches a
maximum of about 80\% as $|\Omega_0|\ne|\Omega_1|\sim\Delta/10$ introduces an
effective detuning for the two-photon transition;  
the adiabatic elimination gives a good approximation since we are still in the
regime of $|\Omega_0|,|\Omega_1|\ll|\Delta|$. 
When the magnitudes of $|\Omega_0|$ and $|\Omega_1|$ are not much smaller than
$|\Delta|$, the population of the excited state is no longer negligible, but
the complete population transfer between states $\ket{0}$ and $\ket{1}$ is
still achievable; see the bottom-left plot in Fig.~\ref{fig:RamanOnRe}. 
The adiabatic elimination method fails in this case of stronger coupling
between the relevant states $\ket{0}$, $\ket{1}$ and the excited state
$\ket{\mathrm{e}}$.

\subsection{Non-zero overall detuning ($\delta\ne0$)}
\label{ssec:New3}
As discussed in Section \ref{ssec:AE3}, one needs to adjust the detuning
$\delta$ to compensate for the light shifts and achieve complete population
transfer from $\ket{0}$ to $\ket{1}$.  
Besides this, there can also be other experimental reasons for choosing a
particular $\delta$ value.
Thus, the situation of $\delta\ne0$ is of practical interest, 
and so we will regard the overall detuning $\delta$ as a free parameter that is
small compared with the average detuning $\Delta$. 
Typically, the ratios $\delta/\Delta$ and $\Omega^{\dagger}\Omega/\Delta^2$
are of the same small order, a few percent perhaps.

The splitting of $H_{\mathrm{I}}^2$ in (\ref{eq:HI2})--(\ref{eq:H_1})
has the diagonal blocks in $M_0^2$, including the $\delta$-dependent
contributions, whereas $\epsilon$ contains the off-diagonal
parts linear in $\delta$. 
We shall treat $\epsilon$ as a small quantity, without, however, regarding
$\delta$ as an expansion parameter. 
Rather, the full $\delta$-dependence of $M_0^2$ is taken into account. 
Then, the eigenvalues of $M^2$ agree with the eigenvalues of $M_0^2$ to first
order in the perturbation of $\epsilon$, and the leading correction will be of
order $\epsilon^2$.

The eigenvalues of $M_0^2$ in the subspace spanned by $\ket{0}$ and $\ket{1}$ 
are
\begin{equation}
\mu_\pm^2=\frac{1}{4}\bigl(\Delta^2+\delta^2\bigr)
+\frac{1}{8}\Omega^{\dagger}\Omega
\pm\frac{1}{8}\sqrt{\bigl(\Omega^{\dagger}\Omega\bigr)^2
+8\delta\Delta\Omega^{\dagger}\sigma_3\Omega+(4\delta\Delta)^2}\,,
\end{equation}
which we get from diagonalizing the upper $2\times2$ diagonal block
$(\Delta+\delta\sigma_3)^2+\Omega\Omega^\dagger$.
The unnormalized eigencolumns are
\begin{equation}
  \left(\begin{array}{c}
      \ds\bigl[4\mu^2-(\Delta+\delta\sigma_3)^2\bigr]\Omega\\0
    \end{array}\right)\quad\mbox{for $\mu_{\ }^2=\mu_{\pm}^2$}\,,
\end{equation}
unless $\Omega$ is an eigencolumn of $\sigma_3$, which is a case of no
interest.  
The eigenvalue 
$\frac{1}{4}\bigl(\Delta^2+\Omega^{\dagger}\Omega\bigr)$ in the
$1\times1$ block for $\ket{\mathrm{e}}$ does not depend on $\delta$. 
With its eigenvalues and eigencolumns at hand, all functions of $M_0^2$ are
readily evaluated.

In passing, we observe that the difference between $\mu_+^2$ and $\mu_-^2$ is
smallest, as a function of $\delta$, when (\ref{delta1}) holds, i.e.,
$4\delta\Delta=-\Omega^{\dagger}\sigma_3\Omega$.
Then
\begin{equation}
  \mu_+^2-\mu_-^2=\frac{1}{4}\sqrt{\bigl(\Omega^{\dagger}\Omega\bigr)^2
-\bigl(\Omega^{\dagger}\sigma_3\Omega\bigr)^2}=\frac{|\Omega_0|\,|\Omega_1|}{2}\,,
\end{equation}
which is nonzero in all situations of interest.

For a systematic inclusion of correction of orders $\epsilon$, $\epsilon^2$,
$\epsilon^3$, \dots, we do not use the perturbation theory for an
approximation of 
the eigenvalues and eigencolumns of $M^2$ for use in (\ref{eq:EvoR}).
Rather, we generate approximations for the evolution matrix 
${U(t)=\exp(-\I H_{\mathrm{I}}t/\hbar)}$ itself with
the aid of an equation of Lippmann-Schwinger type,
\begin{equation}\label{eq:UtExactR}
U(t)=U_0^{\mathrm{(R)}}(t)
-\int_0^t dt'\;\frac{\sin\bigl(M_0(t-t')\bigr)}{M_0}\epsilon U(t')\;,
\end{equation}
where 
\begin{equation}\label{eq:U0R}
U_0^{\mathrm{(R)}}(t)=\cos(M_0t)-\frac{\I}{\hbar}\frac{\sin(M_0t)}{M_0}H_{\mathrm{I}}
\end{equation}
differs from (\ref{eq:EvoR0}) by the inclusion of the $\delta$ dependent terms
in $M_0$ and $H_{\mathrm{I}}$.
One way of verifying that (\ref{eq:UtExactR}) is correct, is by checking that
both sides have the same Laplace transform.
Indeed, they do:
\begin{equation}\label{eq:eqb}
\int_0^\infty dt\,\mathrm{e}^{-st}U(t)=
\frac{1}{s+\I H_{\mathrm{I}}/\hbar}=
\frac{s}{s^2+M_0^2}-\frac{\I}{\hbar}\frac{1}{s^2+M_0^2}H_{\mathrm{I}}
-\frac{1}{s^2+M_0^2}\epsilon\frac{1}{s+\I H_{\mathrm{I}}/\hbar}
\end{equation}
is an identity that follows from (\ref{eq:HI2}).

As mentioned at the end of Section~\ref{ssec:New1}, 
the multiplication order of $H_{\mathrm{I}}$ and even powers of $M^2$ is irrelevant
in (\ref{eq:EvoR}) as they commute. 
In (\ref{eq:U0R}), however, the order does matter 
as $[H_{\mathrm{I}},M_0^2]\neq0$ when $\delta\neq0$.
In addition to (\ref{eq:UtExactR}) with (\ref{eq:U0R}) where  
$H_{\mathrm{I}}$ is on the right, we have, therefore, also an ``on the left''
version, 
\begin{equation}\label{eq:UtExactL}
U(t)=U_0^{(\mathrm{L})}(t)-\int_0^t dt' \;U(t-t')\epsilon\frac{\sin(M_0t')}{M_0}
\end{equation}
with
\begin{equation}\label{eq:U0L}
U_0^{(\mathrm{L})}(t)=\cos(M_0t)
-\frac{\I}{\hbar}H_{\mathrm{I}}\frac{\sin(M_0t)}{M_0}\,.
\end{equation}
Half their sum gives a symmetrized version,
\begin{equation}\label{eq:UtExactS}
U(t)=U_0^{(\mathrm{S})}(t)
-\frac{1}{2}\int_0^t dt'\;\frac{\sin\bigl(M_0(t-t')\bigr)}{M_0}\epsilon U(t')
-\frac{1}{2}\int_0^t dt'\;U(t-t')\epsilon\frac{\sin(M_0t')}{M_0}
\end{equation}
with
\begin{equation}\label{eq:U0S}
U_0^{(\mathrm{S})}(t)=\cos(M_0t)
-\frac{\I}{2\hbar}\frac{\sin(M_0t)}{M_0}H_{\mathrm{I}}
-\frac{\I}{2\hbar}H_{\mathrm{I}}\frac{\sin(M_0t)}{M_0}\,,
\end{equation}
and there are many more variants that one could explore.

Each of the integral equations (\ref{eq:UtExactR}), (\ref{eq:UtExactL}), and
(\ref{eq:UtExactS}) provides a hierarchy of approximations by an iteration
that commences with the respective zeroth-order approximation. 
This is, of course, the procedure by which one generates the Born series from
the Lippmann-Schwinger equation.
For the ``on the right'' equation (\ref{eq:UtExactR}), the
$k$th-order approximation is
\begin{equation}\label{eq:UtkR}
U_k^{\mathrm{(R)}}(t)=U_{k-1}^{\mathrm{(R)}}(t)
-\int_0^t dt'\;\frac{\sin\bigl(M_0(t-t')\bigr)}{M_0}
\epsilon U_{k-1}^{\mathrm{(R)}}(t')\,,
\end{equation}
and analogous expressions apply to the ``on the left'' version and the
symmetrized variant.
Note that $U_k^{\mathrm{(S)}}(t)$ is not half the sum of
$U_k^{\mathrm{(R)}}(t)$ and $U_k^{\mathrm{(L)}}(t)$ for $k\ne0$; that arithmetic mean
could also be taken as a valid $k$th-order approximation.
For such a scheme to be useful in practice, the zeroth-order approximation
should be quite good to begin with, the first-order approximation should be
sufficient for many purposes, and it should not be necessary to go beyond the
second order. 
 
The various $k$th-order approximations differ from each other, but they are all
accurate up to $k$th-order in
$\bigl|\epsilon/M_0^2\bigr|\sim|\delta\Omega|/\Delta^2$. 
In the common case where $|\Omega/\Delta|\sim10^{-1}$ and 
$|\delta/\Delta|\sim10^{-2}$, we have $|\epsilon/M_0^2|\sim10^{-3}$. 
Remember that the $\delta$ dependence in $H_{\mathrm{I}}$ and the $\epsilon$
dependence (which also depends on $\delta$) are treated separately. 
Although $\epsilon$ goes to zero when $\delta$ vanishes, 
$|\epsilon/{M_0}^2|$ is one order of magnitude smaller than $|\delta/\Delta|$.

A technical point is the following.
The approximate evolution matrices $U_k(t)$ are not unitary, rather 
$U_k(t)^{\dagger}U_k(t)$ deviates from the unit matrix by an amount of order 
$\epsilon^{k+1}$.
One can cope with this in various ways~\cite{HanThesis}.
Perhaps the simplest is to ensure proper normalization by including a
time-dependent factor that depends on the initial set of probability
amplitudes, thereby arriving at an effectively unitary matrix
$\widetilde{U}_k(t)$ that is 
suitable for the given initial column $\Psi_{\mathrm{I}}(0)$,
\begin{equation}\label{eq:widetildeU}
\widetilde{U}_k(t)\Psi_{\mathrm{I}}(0)
=\frac{U_k(t)\Psi_{\mathrm{I}}(0)}
{\sqrt{\Psi_{\mathrm{I}}(0)^{\dagger}U_k(t)^{\dagger}U_k(t)\Psi_{\mathrm{I}}(0)}}\,.
\end{equation}
In other words, we apply $U_k(t)$ to $\Psi_{\mathrm{I}}(0)$ and normalize the
resulting column to unit length. 
This procedure worked fine for all examples that we studied. 

As remarked above, the $k$th-order approximations $U_k^{\mathrm{(R)}}(t)$,
$U_k^{\mathrm{(L)}}(t)$, and $U_k^{\mathrm{(S)}}(t)$ differ slightly and might
not describe the system equally well. 
We discuss a few examples of the state populations as functions of time under
the different approximations, and analyze their performance.
We compare the results to the exact numerical answers. 
The value $\Delta=400\,\mathrm{MHz}$ is taken for the average detuning, as it
is of typical order for real experiments, and the values of the Rabi coupling
strengths 
$\Omega_0$ and $\Omega_1$ can vary in a range of fractions of $\Delta$. 
The overall detuning $\delta$ of the two-photon transition can be controlled
to within $1\,\mathrm{MHz}$ accuracy in laboratory experiments. 
For the purpose of this analysis, then, we take the liberty of setting
$\delta$ to any value we like.

\begin{figure}
\centering
\center{
\begin{picture}(400,205)(0,-7)
\put(10,0){\includegraphics[scale=1]{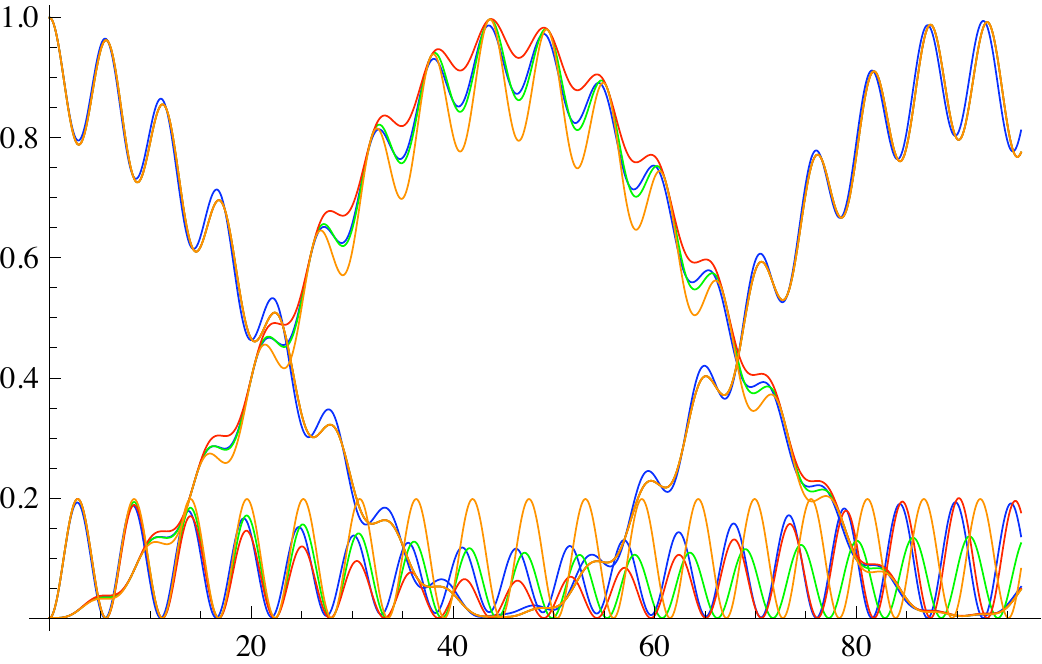}\label{subfig:OffR7}}
\put(230,-5){\makebox(0,0)[b]{$\Delta t$}}
\put(340,130){\linethickness{0.8pt}\color{blue}\line(1,0){20}}
\put(380,130){\makebox(0,0){Exact}}
\put(340,110){\linethickness{0.8pt}\color{red}\line(1,0){20}}
\put(380,110){\makebox(0,0){$\widetilde{U}_0^{\mathrm{(R)}}(t)$}}
\put(340,90){\linethickness{0.8pt}\color{green}\line(1,0){20}}
\put(380,90){\makebox(0,0){$\widetilde{U}_0^{\mathrm{(S)}}(t)$}}
\put(340,70){\linethickness{0.8pt}\color{orange}\line(1,0){20}}
\put(380,70){\makebox(0,0){$\widetilde{U}_0^{\mathrm{(L)}}(t)$}}
\end{picture}}
\caption{\label{fig:Off1}
  Plots of populations obtained from different zeroth-order
  solutions of the Lippmann-Schwinger equations. 
  Blue curves give the exact results from numerical simulation 
  using the Hamiltonian $H_{\mathrm{I}}$; 
  green curves show solutions from the symmetric approximation
  $\widetilde{U}_0^{\mathrm{(S)}}(t)$; 
  red curves show solutions from $\widetilde{U}_0^{\mathrm{(R)}}(t)$; 
  and orange curves show solutions from $\widetilde{U}_0^{\mathrm{(L)}}(t)$. 
  The parameters are $\Delta=400\,\mathrm{MHz}$,
  $|\Omega_0|=\Delta/2$, $|\Omega_1|=3\Delta/10$ and
  $\delta=-\Omega^{\dagger}\sigma_3\Omega/(4\Delta)
   =\bigl(|\Omega_1|^2-|\Omega_0|^2\bigr)/(4\Delta)=-16\,\mathrm{MHz}$.
  The effective Rabi frequency is $\Omega_{\mathrm{R}}=27.8\,\mathrm{MHz}$,
  about 7\% of $\Delta$. 
  Initially, we have $c_0(0)=1$ and $c_1(0)=c_{\mathrm{e}}(0)=0$.
  The curves starting at $1$ show the approximations for
  $\bigl|c_0(t)\bigr|^2$; the curves that start at $0$ and rise to $1$
  are for $\bigl|c_1(t)\bigr|^2$; and the curves that start at $0$ and never
  exceed small values are for $\bigl|c_{\mathrm{e}}(t)\bigr|^2$. }
\end{figure}
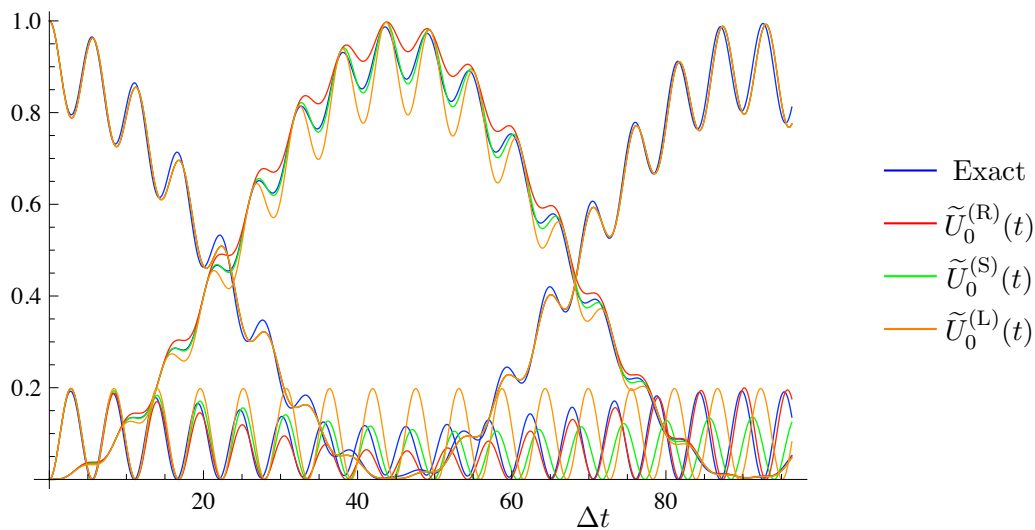

Figure \ref{fig:Off1} gives an example that demonstrates the quality of the
zeroth-order approximations given by $\widetilde{U}_0^{\mathrm{(R)}}(t)$,
 $\widetilde{U}_0^{\mathrm{(L)}}(t)$, and $\widetilde{U}_0^{\mathrm{(S)}}(t)$.
Since $U_0^{\mathrm{(S)}}(t)$ is an average of $U_0^{\mathrm{(R)}}(t)$ and
$U_0^{\mathrm{(L)}}(t)$,  
we expect the population curves for $\widetilde{U}_0^{\mathrm{(S)}}(t)$ to lie
between the other two curves and this can be seen quite clearly in
Fig.~\ref{fig:Off1}.  
Moreover, the solutions of the population in the initial state $\ket{0}$ do
not depend much on which of the three zeroth-order approximations is used, 
and all of them are very close to the exact numerical solution. 
In this example, we use $\delta=(|\Omega_1|^2-|\Omega_0|^2)/(4\Delta)$, the
value of (\ref{delta1}), which gives an effective resonant two-photon
transition, and we can see that full population transfer from $\ket{0}$ to
$\ket{1}$ can be achieved.  
In comparison, full population transfer cannot be achieved for other values of
$\delta$. 

In Fig.~\ref{fig:Off1}, the value of $|\delta|$ is about a few percent of
$|\Delta|$. 
The plot shows that, in this parameter regime,
$\widetilde{U}_0^{\mathrm{(S)}}(t)$ approximates the evolution of the
probabilities for finding $\ket{0}$ and $\ket{1}$ quite well, and it certainly
works best among the three different zeroth-order approximations shown here. 
The approximation for the excited state population also works well when $t$ is
short, i.e., during the first few fast oscillation periods, but the deviation
grows quickly with time.

\begin{figure}
{\centering
\begin{picture}(456,280)(-1,0)
\put(0,270){\makebox(0,0)[l]{(a)}}
\put(0,30){\includegraphics[scale=0.73]{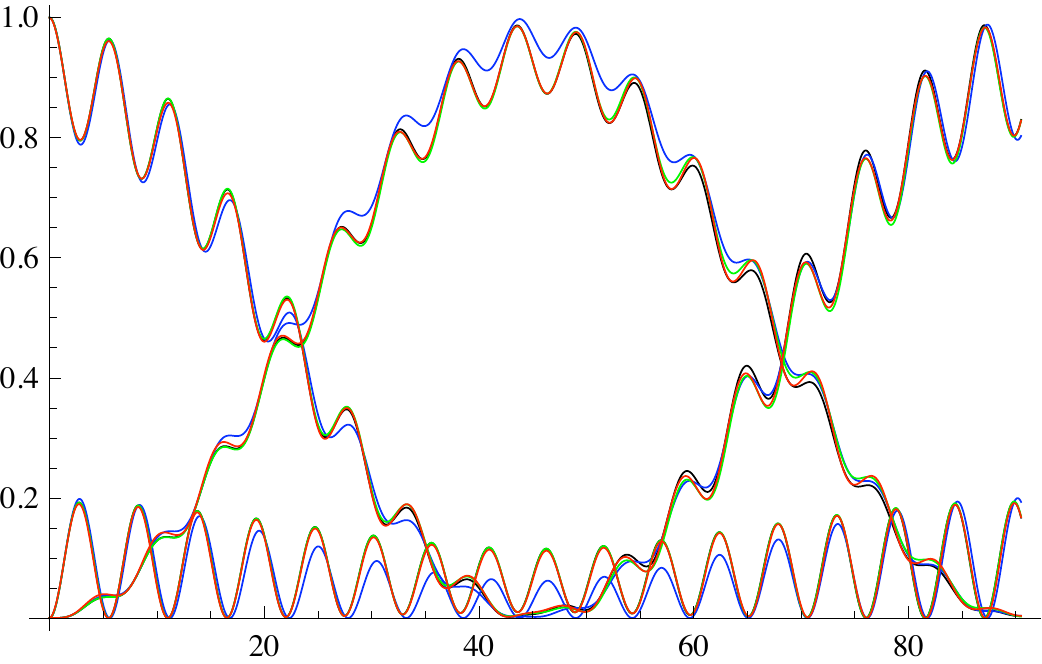}}
\put(25,0){\put(3,180){\includegraphics[scale=0.21]{OffRB1zoom.pdf}}
\put(2,180){\color{gray}\framebox(110,90){ }}
\put(118,179){\includegraphics[scale=0.35]{OffRB2zoom.pdf}}
\put(117,180){\color{gray}\framebox(80,90){ }}
}
\put(96,140){\color{gray}\framebox(38,30){ }}
\put(115,171){\linethickness{0.8pt}\color{gray}\vector(-1,2){5}}
\put(150,83){\color{gray}\framebox(30,35){ }}
\put(165,118){\linethickness{0.8pt}\color{gray}\vector(0,1){62}}
\put(215,32){\makebox(0,0){$\Delta t$}}
\put(25,15){\makebox(0,0){Exact}}
\put(15,5){\linethickness{0.8pt}\color{black}\line(1,0){20}}
\put(70,5){\linethickness{0.8pt}\color{blue}\line(1,0){20}}
\put(80,15){\makebox(0,0){$\widetilde{U}_0^{\mathrm{(R)}}(t)$}}
\put(125,5){\linethickness{0.8pt}\color{green}\line(1,0){20}}
\put(135,15){\makebox(0,0){$\widetilde{U}_1^{\mathrm{(R)}}(t)$}}
\put(180,5){\linethickness{0.8pt}\color{red}\line(1,0){20}}
\put(190,15){\makebox(0,0){$\widetilde{U}_2^{\mathrm{(R)}}(t)$}}
\put(10,0){
\put(225,270){\makebox(0,0)[l]{(b)}}
\put(225,30){\includegraphics[scale=0.73]{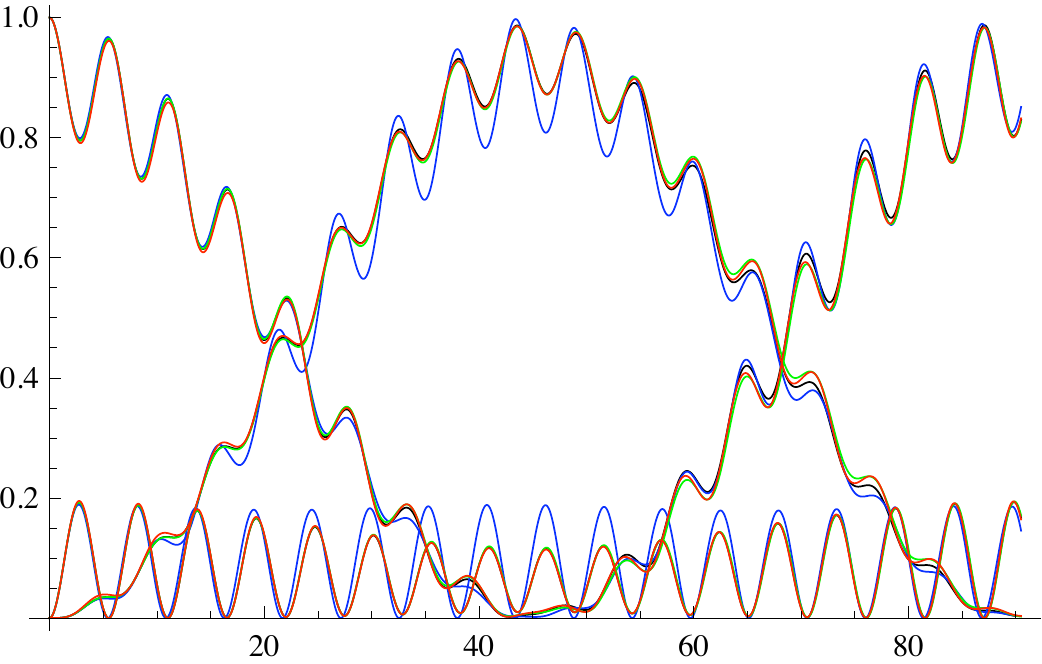}}
\put(435,32){\makebox(0,0){$\Delta t$}}
\put(5,0){
\put(243,180){\includegraphics[width=103\unitlength,height=90\unitlength]
{OffRF1zoom.pdf}}
\put(243,180){\color{gray}\framebox(103,90){ }}
\put(352,181){\includegraphics[width=75\unitlength,height=90\unitlength]
{OffRF2zoom.pdf}}
\put(351,180){\color{gray}\framebox(75,90){ }}
}
\put(322,136){\color{gray}\framebox(38,30){ }}
\put(340,166){\linethickness{0.8pt}\color{gray}\vector(-1,2){7}}
\put(375,83){\color{gray}\framebox(30,37){ }}
\put(390,120){\linethickness{0.8pt}\color{gray}\vector(0,1){60}}
\put(260,15){\makebox(0,0){Exact}}
\put(250,5){\linethickness{0.8pt}\color{black}\line(1,0){20}}
\put(305,5){\linethickness{0.8pt}\color{blue}\line(1,0){20}}
\put(315,15){\makebox(0,0){$\widetilde{U}_0^{\mathrm{(L)}}(t)$}}
\put(360,5){\linethickness{0.8pt}\color{green}\line(1,0){20}}
\put(370,15){\makebox(0,0){$\widetilde{U}_1^{\mathrm{(L)}}(t)$}}
\put(415,5){\linethickness{0.8pt}\color{red}\line(1,0){20}}
\put(425,15){\makebox(0,0){$\widetilde{U}_2^{\mathrm{(L)}}(t)$}}
}
\end{picture}}
\caption{\label{fig:OffRB}
  Comparison  of the probabilities obtained 
  from the zeroth-, first-, and second-order
  approximations of the Lippmann-Schwinger equations (\ref{eq:UtExactR})
  and (\ref{eq:UtExactL}). 
  In plot (a), the blue, green, and red curves show solutions from
  $\widetilde{U}_0^{\mathrm{(R)}}(t)$, $\widetilde{U}_1^{\mathrm{(R)}}(t)$, 
  and $\widetilde{U}_2^{\mathrm{(R)}}(t)$, respectively; 
  and in plot (b), the blue, green, and red curves show
  solutions from from
  $\widetilde{U}_0^{\mathrm{(L)}}(t)$, $\widetilde{U}_1^{\mathrm{(L)}}(t)$, 
  and $\widetilde{U}_2^{\mathrm{(L)}}(t)$, respectively.
  The parameters and the initial state are the same as in Fig.~\ref{fig:Off1}.}
\end{figure}

\begin{figure}
{\centering
\begin{picture}(440,230)(0,0)
\put(10,20){\includegraphics[scale=1]{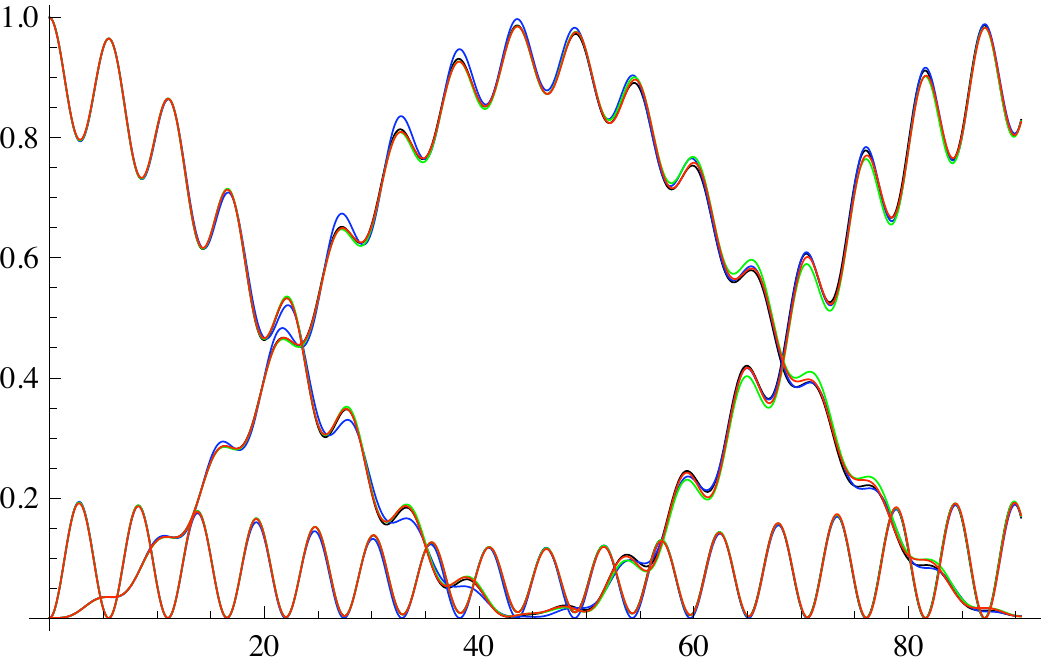}}
\put(290,25){\makebox(0,0){$\Delta t$}}
\put(150,10){\makebox(0,0){Exact}}
\put(140,0){\linethickness{0.8pt}\color{black}\line(1,0){20}}
\put(200,0){\linethickness{0.8pt}\color{blue}\line(1,0){20}}
\put(210,10){\makebox(0,0){$\widetilde{U}_0^{\mathrm{(S)}}(t)$}}
\put(260,0){\linethickness{0.8pt}\color{green}\line(1,0){20}}
\put(270,10){\makebox(0,0){$\widetilde{U}_1^{\mathrm{(S)}}(t)$}}
\put(310,0){\linethickness{0.8pt}\color{red}\line(1,0){20}}
\put(320,10){\makebox(0,0){$\widetilde{U}_2^{\mathrm{(S)}}(t)$}}
\put(320,142){\includegraphics[scale=0.5]{OffRS1zoom.pdf}}
\put(148,175){\color{gray}\framebox(40,32){ }}
\put(320,140){\color{gray}\framebox(116,95){ }}
 \put(188,190){\linethickness{0.8pt}\color{gray}\vector(1,0){132}}
  \put(351,22){\includegraphics[scale=0.4]{OffRS2zoom.pdf}}
 \put(220,93){\color{gray}\framebox(32,48){ }}
\put(350,22){\color{gray}\framebox(75,108){ }}
 \put(252,118){\linethickness{0.8pt}\color{gray}\vector(3,-1){96}}
\end{picture}}
\caption{\label{fig:OffRS}
  Comparison of the zeroth-, first-, and second-order approximations of the
  symmetric Lippmann-Schwinger (\ref{eq:UtExactS}).
  The parameter values, the initial state, and the color coding are the same
  as in Fig.~\ref{fig:OffRB}.}
\end{figure}

The accuracy is better for higher-order approximations. 
We compare the approximations of zeroth, first, and second order 
for the three different Lippmann-Schwinger equations in
Figs.~\ref{fig:OffRB}(a), \ref{fig:OffRB}(b), and \ref{fig:OffRS}. 
Figure \ref{fig:OffRB}(a) shows that the deviation of the zeroth-order
approximation $\widetilde{U}_0^{\mathrm{(R)}}(t)$ from the exact numerical
solution is large when about half of the effective Rabi cycle is completed,
i.e., around $t=\pi/\Omega_{\mathrm{R}}$ or $\Delta t\simeq45$, 
and the deviation is smaller around a full Rabi cycle. 
The main deviation is in the size of the small-amplitude oscillations
with short period, whereas the Rabi oscillation with longer period  
is reproduced equally well by $\widetilde{U}_k^{\mathrm{(R)}}(t)$ with $k=0,1,2$.  
The first-order approximation corrects part of the error in the zeroth-order
approximation, and the second-order approximation improves matters further and
gets the probabilities very close to their exact values. 
The same observations can be made about the
corresponding ``on the left'' approximations in Fig.~\ref{fig:OffRB}(b). 
How about the symmetric version $\widetilde{U}_k^{\mathrm{(S)}}(t)$ whose
zeroth-order approximation already works quite well? 

Figure \ref{fig:OffRS} shows that when the two-photon transition is resonant,
the difference between $\widetilde{U}_0^{\mathrm{(S)}}(t)$,
$\widetilde{U}_1^{\mathrm{(S)}}(t)$, and $\widetilde{U}_2^{\mathrm{(S)}}(t)$
is difficult to detect. 
All three lowest-order approximations of (\ref{eq:UtExactS}) describe the
evolution of the system well during the first Rabi cycle. 
The zeroth-order approximation works surprisingly well, at times it gives a
better result than the higher-order approximations (see the bottom plot of the
two blow-ups). 
The improvement offered by the higher-order approximations 
can be observed near the middle of the Rabi cycle (see the top plot of the
blow-ups), and this improvement is more substantial when $\epsilon$ gets
larger. 
For a practical application to experiments that aim at complete population
transfer from $\ket{0}$ to $\ket{1}$, the approximation provided by
$\widetilde{U}_0^{\mathrm{(S)}}(t)$ is accurate enough to determine the
parameter values reliably.

\subsection{Discussion}\label{ssec:New4}
To summarize our approach, we split $(H_{\mathrm{I}}/\hbar)^2$, the square of
the interaction Hamiltonian, into two parts: 
the dominant part $M_0^2$ and a small correction $\epsilon$. 
With this splitting and any one of the Lippmann-Schwinger equations
(\ref{eq:UtExactR}), (\ref{eq:UtExactL}), or (\ref{eq:UtExactS}), successive
approximations to the unitary evolution matrix 
$U(t)=\exp(-\I H_{\mathrm{I}}t/\hbar)$ can be obtained iteratively. 
If we use the approximations given by $\widetilde{U}_k^{\mathrm{(R)}}(t)$ 
and $\widetilde{U}_k^{\mathrm{(L)}}(t)$, 
we only need to do one iteration and stop at the first-order solution
(${k=1}$) for a very good approximation. 
If we use the symmetric version $\widetilde{U}_k^{\mathrm{(S)}}(t)$, 
we do not even need to go beyond the zeroth-order approximation since 
$\widetilde{U}_0^{\mathrm{(S)}}(t)$ is already very close to the exact
evolution for typical experimental parameters. 
Thus, we have
\begin{equation}\label{eq:UtAppro}
U(t)\simeq
U_0^{\mathrm{(S)}}(t)=\cos(M_0t)
-\frac{\I}{2\hbar}\frac{\sin(M_0t)}{M_0}H_{\mathrm{I}}
-\frac{\I}{2\hbar}H_{\mathrm{I}}\frac{\sin(M_0t)}{M_0}\to
\widetilde{U}_0^{\mathrm{(S)}}(t)\,,
\end{equation}
where $\widetilde{U}_0^{\mathrm{(S)}}(t)$ differs from $U_0^{\mathrm{(S)}}(t)$
by the time-dependent factor of (\ref{eq:widetildeU}) that ensures 
$\Psi_{\mathrm{I}}(t)=\widetilde{U}_0^{\mathrm{(S)}}(t)\Psi_{\mathrm{I}}(0)$
is properly normalized for the given initial column of probability amplitudes.

According to (\ref{eq:UtAppro}), the oscillation of the state populations of
$\ket{0}$ and $\ket{1}$ are governed by the operator $M_0^2$. 
The effective Rabi oscillation frequency, to a very good approximation, only
depends on the eigenvalues $\mu_\pm^2$ of the first diagonal block of
$M_0^2$. 
Applying the same argument as in Section \ref{ssec:New2} for the case of 
$\delta=-\Omega^\dagger\sigma_3\Omega/4\Delta$ in (\ref{delta1}), the
effective Rabi frequency is (we take $\mu_\pm>0$)  
\begin{equation}\label{OmegaR}
\Omega_{\mathrm{R}}=\mu_+-\mu_-
=\frac{1}{2}\sqrt{(\Delta^2+\delta^2)+\frac{1}{2}(|\Omega_0|+|\Omega_1|)^2}
-\frac{1}{2}\sqrt{(\Delta^2+\delta^2)+\frac{1}{2}(|\Omega_0|-|\Omega_1|)^2}\,.
\end{equation}
With this particular choice of $\delta$, the Rabi oscillation amplitude could
reach unity;  
however, this might not be the real maximum that the population in state
$\ket{1}$ can 
reach, because on top of this slow effective Rabi oscillation, the population
also oscillates with a fast frequency. 
This fast oscillation goes roughly with the frequency
$\mu_++\mu_-\simeq|\Delta|$. 
Nevertheless, up to linear order in $\delta/\Delta$, we find that the
oscillation amplitude goes to unity, regardless whether we 
choose $\delta$ in accordance with (\ref{delta1}) or with
(\ref{delta2}).
This shows why the evolution of the system is essentially the same for both
$\delta$ values. 

\begin{figure}
\centering
{\begin{picture}(410,205)(-5,-8)
\put(0,0){\includegraphics[scale=0.85]{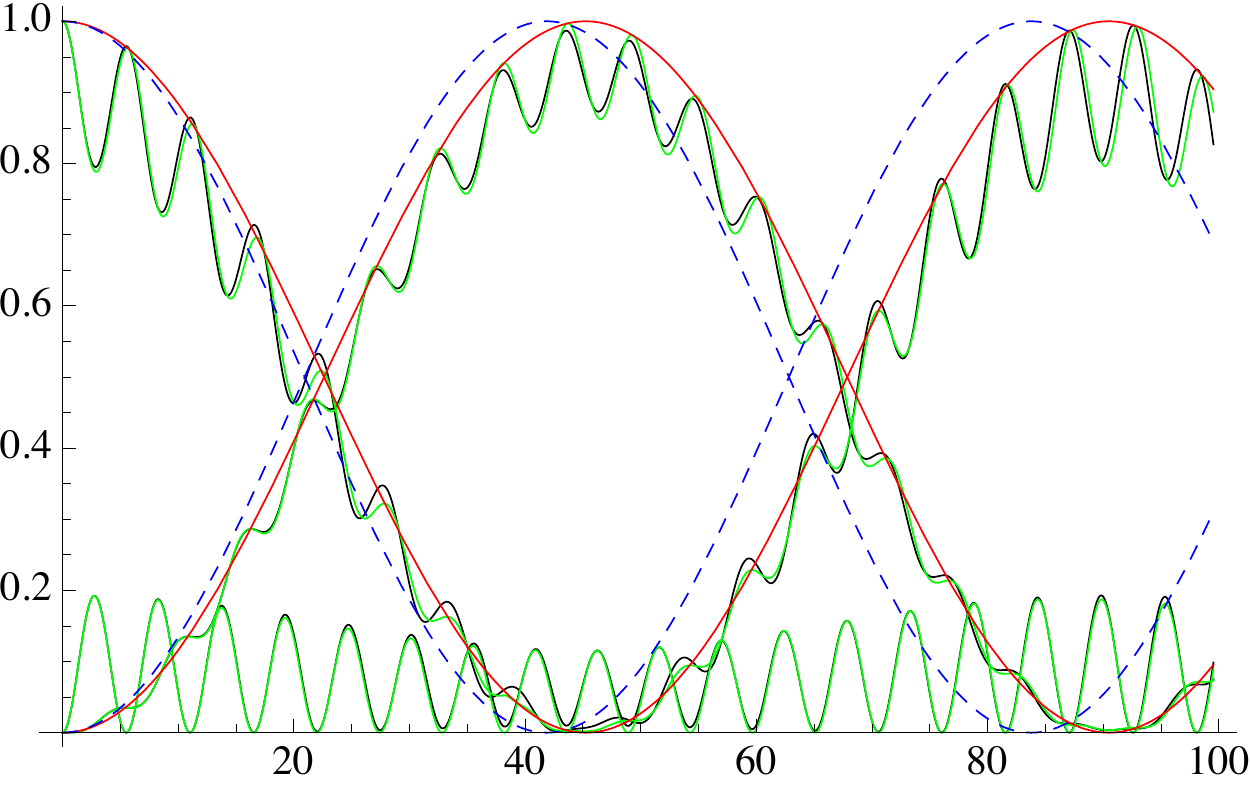}}
\put(267,0){\makebox(0,0){$\Delta t$}}
\put(320,120){\linethickness{0.8pt}\color{black}\line(1,0){20}}
\put(360,120){\makebox(0,0){Exact}}
\put(320,100){\linethickness{0.8pt}\color{green}\line(1,0){20}}
\put(360,100){\makebox(0,0){$\widetilde{U}_0^{\mathrm{(S)}}(t)$}}
\put(320,72){{\color{blue}{- - -}}}
\put(370,80){\makebox(0,0){Adiabatic}}
\put(370,70){\makebox(0,0){elimination}}
\put(320,50){\linethickness{0.8pt}\color{red}\line(1,0){20}}
\put(360,50){\makebox(0,0){$\E^{-\I M_0t}$}}
\end{picture}}
\caption{\label{fig:M0}
  Improvement on the effective two-level Hamiltonian compared with
  adiabatic elimination. 
  The black curves give the exact numerical solution; 
  the green curves are for the symmetric zeroth-order approximation of
  $U(t)$; 
  the blue dashed curves are for the adiabatic-elimination approximation; 
  and the red curves result from taking $M_0$ as the effective Hamiltonian. 
  The parameters are the same as in Fig.~\ref{fig:Off1}.}
\end{figure}
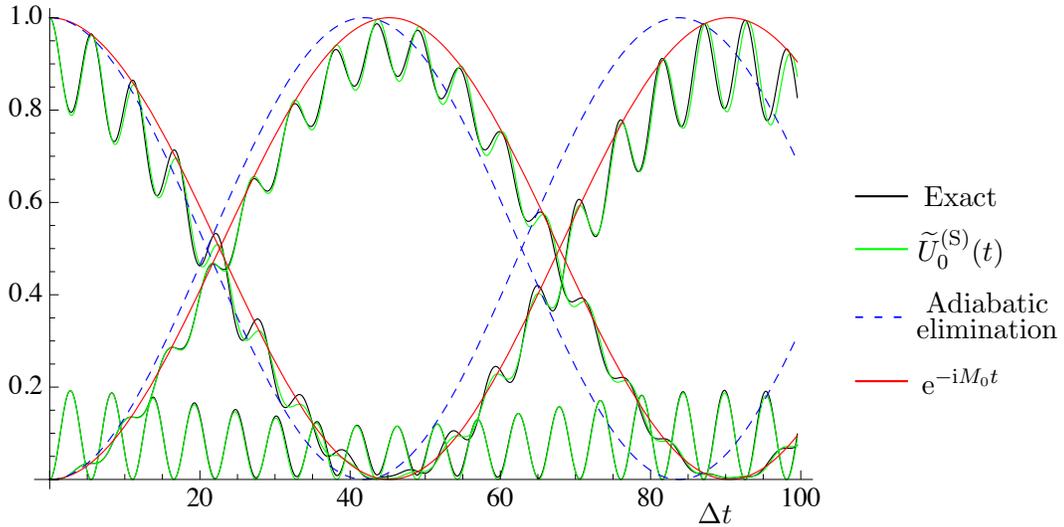

Moreover, since the effective Rabi oscillation of the two relevant states
depends on the eigensystem of $M_0^2$ only, the effective Hamiltonian between
states $\ket{0}$ and $\ket{1}$ is 
approximately given by the $2\times2$ upper diagonal block of $\hbar M_0$, if
we exclude the excited state $\ket{\mathrm{e}}$ from the evolution directly.  
That is
\begin{equation}
H_{\mathrm{eff}}=-\frac{\hbar}{2}\sqrt{(\Delta+\delta\sigma_3)^2
+\Omega\Omega^\dagger}\,.
\end{equation}
The minus sign is chosen with the same reasoning as in~(\ref{M_00}). From this
effective Hamiltonian, we can find the Rabi frequency directly and one of the
special cases was already given in~(\ref{OmegaR}); the oscillation amplitude
for an arbitrary $\delta$ is 
\begin{equation}
P=1-\frac{(\Omega^\dagger\sigma_3\Omega+4\delta\Delta)^2}{(\Omega^\dagger\Omega)^2
+8\delta\Delta\Omega^\dagger\sigma_3\Omega+(4\delta\Delta)^2}\,,
\end{equation}
and again we have $P=1$ when $\delta=-\Omega^\dagger\sigma_3\Omega/4\Delta$. 
Figure~\ref{fig:M0} shows that this effective Hamiltonian is much
more accurate than that of the adiabatic-elimination approximation, 
inasmuch as the evolution by $\E^{-\I M_0t}$ gives a very close envelope
of the population oscillation.

\section{Summary and Outlook}
After reviewing the standard adiabatic-elimination approximation, which
reduces the theoretical description of a three-level Raman transition to an
effective two-level problem, and identifying some of the shortcomings of this
approach, we introduced an alternative approximation method.
Similarly to adiabatic elimination, there is an essential two-level component
in the new method without, however, eliminating the third auxiliary level.
This makes the new method easy to use, inasmuch as one only needs to
diagonalize a $2\times2$ matrix.
Integro-differential equations of Lippmann-Schwinger type are the powerful
tools that enable us to generate successive approximations.
A particular one with high symmetry performs so well that the lowest-order
approximation is all one needs for a highly reliable determination of
experimental parameters.

For the sake of simplicity in presentation, the discussion was here limited to
two-photon transitions.
The method can also be applied to more complicated situations
\cite{HanThesis}, about which we will report on another occasion.

\section*{Acknowledgments}
This work is supported by the National Research Foundation and the Ministry of
Education, Singapore. We are grateful for insightful discussions with Vanessa
Paulisch, Antoine Browaeys, J\'anos Bergou, Gerd Leuchs and Bj\"orn Hessmo.

\end{document}